\documentclass[letterpaper,twocolumn,10pt]{article}
\usepackage{usenix2019_v3}

\usepackage{algorithm}
\usepackage{amsthm}
\usepackage{amssymb}
\usepackage{multirow}
\usepackage{tabls}
\usepackage{bigstrut}
\usepackage{comment}
\usepackage[noend]{algpseudocode}

\usepackage{subcaption}
\usepackage{textcomp}
\usepackage{tabularx}
\usepackage{tabu}
\usepackage{longtable}
\usepackage{enumitem}
\usepackage{makecell}
\usepackage{multirow}
\usepackage{xspace}
\usepackage{url}
\usepackage{flushend}
\usepackage{booktabs}
\usepackage{graphicx}
\makeatletter
\def\do@url@hyp{\do\-\do\_}
\makeatother

\newcommand{\PHB}[1]{\noindent\textbf{#1}} 
\newcommand{\PHM}[1]{\noindent\textbf{#1}} 

\newcommand{\SysName}{\texttt{Torpor}\xspace}
\newcommand{\Cloud}{Alibaba Cloud\xspace}

\newcommand{\revise}[1]{{#1}}
\newcommand{\add}[1]{{#1}}
\newcommand{\new}[1]{{#1}}
\newcommand{\shep}[1]{{#1}}

\newcommand{\circledNum}[1]{\raisebox{.5pt}{\textcircled{\raisebox{-.9pt} {#1}}}}

\graphicspath{{figures/}}

\begin{document}

\title{\Large \bf Torpor: GPU-Enabled Serverless Computing for Low-Latency,\\Resource-Efficient Inference}

\date{}


\author{
	{\rm Minchen Yu$^{\dag}$$^{\ddag}$ \quad Ao Wang$^{\S}$ \quad Dong Chen$^{\ddag}$ \quad Haoxuan Yu$^{\ddag}$ \quad Xiaonan Luo$^{\ddag}$ \quad Zhuohao Li$^{\ddag}$}\\
  {\rm Wei Wang$^{\ddag}$ \quad Ruichuan Chen$^{\ast}$ \quad Dapeng Nie$^{\S}$ \quad Haoran Yang$^{\S}$ \quad Yu Ding$^{\S}$}\\
	$^{\dag}$CUHK-Shenzhen \quad  $^{\ddag}$HKUST \quad $^{\S}$Alibaba Group \quad $^{\ast}$Nokia Bell Labs 
} 

\maketitle

\begin{abstract}
    



Serverless computing offers a compelling cloud model for online inference
services. However, existing serverless platforms lack efficient support for
GPUs, hindering their ability to deliver high-performance inference. In this
paper, we present \SysName, a serverless platform for GPU-efficient,
low-latency inference. To enable efficient sharing of a node's GPUs among
numerous inference functions, \SysName maintains models in main memory and
dynamically swaps them onto GPUs upon request arrivals (i.e., late binding
with model swapping). \SysName uses various techniques, including
asynchronous API redirection, GPU runtime sharing, pipelined model execution,
and efficient GPU memory management, to minimize latency overhead caused by
model swapping. Additionally, we design an interference-aware request
scheduling algorithm that utilizes high-speed GPU interconnects to meet
latency service-level objectives (SLOs) for individual inference functions.
We have implemented \SysName and evaluated its
performance in a production environment. Utilizing late binding and model
swapping, \SysName can concurrently serve hundreds of inference functions on
a worker node with 4 GPUs, while achieving latency performance
comparable to native execution, where each model is cached exclusively on a
GPU. 
Pilot deployment in a leading commercial serverless cloud shows that \SysName reduces the GPU provisioning cost by 70\% and 65\% for users and the platform, respectively.

\end{abstract}

\section{Introduction}
\label{sec:intro}

The remarkable advances in machine learning (ML) and its widespread adoption
in various domains have fueled a surging demand for cloud-based ML inference
services~\cite{zhang_caerus_nodate,zhang_mark:_2019,shen_nexus_2019,choi_serving_2022,gujarati_serving_2020}. Severless computing offers a compelling cloud model for inference
serving~\cite{yang_infless_2022, yu_gillis_icdcs, ali_batch_nodate}. In a
serverless cloud, users publish ML models as inference functions, and
delegate resource provisioning and scaling responsibilities to the cloud platform.
Serverless computing is also economically appealing as users only pay for the
resources consumed by their functions (i.e., pay-per-use billing),
eliminating resource idling costs.

However, today's serverless computing platforms, such as AWS Lambda~\cite
{aws_lambda} and Alibaba Function Compute~\cite{ali_fc}, lack efficient
support for GPUs. They typically run an ML model in a container (or a
microVM) and early bind it to a GPU before starting to serve requests. To avoid
considerable startup overhead of on-demand GPU function provisioning (e.g., tens of seconds as shown in Table~\ref
{tab:runtime_isolation}), an inference function is maintained as a long-lived, provisioned
instance on a designated GPU to handle future requests~\cite
{aws_provisioned,fc_gpu}. This approach essentially follows the ``serverful''
inference serving practice~\cite{shen_nexus_2019, romero_infaas_nodate,
zhang_mark:_2019}, requiring users to pay for the occupied GPUs even during
function idling. Furthermore, our analysis in a production cloud demonstrates that
inference functions exhibit varying request rates, with 85\% functions
being invoked no more than once per minute (Fig.~\ref
{fig:motiv_low_rate}). Early binding these functions to GPUs results in low
utilization and imbalanced load across GPUs, making it inefficient for cloud
providers.




We believe that an efficient serverless inference platform should provide
four desirable properties. First, it should enable \emph
{pay-per-GPU-use billing} for users, with charges incurred only when the
functions are invoked and running on GPUs. 
Second, the platform should achieve optimal
GPU utilization through efficient \emph{GPU sharing} for concurrent
inference functions, minimizing resource provisioning costs for cloud providers.
Third, the
platform should be aware of the user-specified \emph{latency SLOs} and strive to
meet them for all inference requests, if feasible. 
Lastly, the platform should achieve the aforementioned three properties \emph{without requiring detailed knowledge about inference models} due to intellectual property and business-critical confidentiality reasons.  We notice that there have been several relevant systems developed in recent years~\cite{choi_memory_nodate,huang_swapadvisor_2020,rhu_vdnn_2016,yu_salus_nodate,yang_infless_2022,shen_nexus_2019,romero_infaas_nodate,deepplan_2023, gujarati_serving_2020}, none of which, however, provide all of these properties for serverless inference. They often suffer from cost inefficiency, SLO violations, or necessitate model-specific knowledge (see \S\ref{sec:background-limitations} and \S\ref{sec:discuss}).

\add{
In this paper, we present \SysName, a GPU-efficient serverless inference
platform that achieves all four desirable properties and is readily-deployable onto real-world serverless platforms without intrusive changes. 
\SysName follows a late binding design principle, whereby idle inference models are maintained in host memory and dynamically swapped to GPUs upon request arrivals. Compared
to GPU memory, host memory is less expensive and has a much larger capacity,
making it an ideal storage for holding numerous idle functions. This approach
naturally supports pay-per-GPU-use billing, as idle functions no longer
occupy GPU resources. Furthermore, by dynamically swapping models from host
to GPUs, it enables fine-grained GPU sharing among concurrent inference
functions, substantially improving GPU utilization and load balancing across
GPUs. Model swapping can also
be efficiently performed through pipelined loading, yielding
significantly lower latency compared to function cold starts.  All these are achieved  without detailed knowledge about inference models -- a must-have in a commercial environment for intellectual property and confidentiality protection. These
techniques, combined with intelligent request scheduling, enable the platform
to optimize the SLO attainment for users.
}

\add{
Specifically, to realize late binding while being readily-deployable on real-world serverless platforms, \SysName leverages a GPU pooling
architecture. In this design, each worker node manages a pool of local GPUs and allows
its inference functions to access any of these GPUs freely through CUDA API
redirection. This enables seamless model swapping within a GPU pool and is
transparent to users. However, this approach also presents three key challenges.
}

\add{
First, GPU pooling and model swapping incur high communication overhead
compared to native execution (i.e., executing a model directly on a GPU).
To address this challenge, \SysName proposes
\emph{asynchronous API redirection} to avoid frequent synchronizations between the
inference functions and the GPU pool, eliminating the high
communication overheads for model inference. \SysName further utilizes
\emph{pipeline execution} to overlap the host-to-GPU model swapping and the
inference execution, thereby reducing
end-to-end latency. It also utilizes high-speed NVLink
for fast model swapping between GPUs whenever feasible and beneficial. 
Combined with
low-latency API redirection, \SysName can efficiently execute models on any
available GPUs. 
\SysName is intentionally designed to be model-agnostic to meet the confidentiality requirements while being generally applicable to various models, including even large generative models where runtime states (e.g., KV cache) can be managed as part of the model.
}

\add{
The second challenge is that GPU pooling and model swapping necessitate an
efficient GPU memory management system. \SysName designs such a system that
automatically tracks the addresses of models as they are swapped across
multiple GPUs and adjusts each memory access of CUDA APIs accordingly during
inference execution. It also efficiently organizes and shares memory blocks
to avoid high memory allocation overheads, improving the overall performance
of model swapping. Additionally, \SysName offers two GPU runtime management modes---runtime sharing and runtime isolation---to meet various needs for resource efficiency and cross-model isolation.
}



The third challenge is that the platform should meet the latency SLOs for inference functions while
maintaining low GPU costs. \SysName proposes three policies to achieve this objective.
First, \SysName designs a request scheduling algorithm that minimizes model
swapping overheads, resulting in reduced end-to-end inference latency. It
categorizes models into two groups, heavy or light, based on whether these
models incur high overhead during swapping via PCIe. \SysName then prioritizes NVLink over PCIe for transferring heavy
models across GPUs, effectively reducing concurrent PCIe traffic. 
Second, \SysName globally manages GPU memory in the pool and leverages model heaviness to
guide eviction. 
Together with request scheduling, this
approach significantly minimizes model swapping overhead. Third, \SysName
proposes an SLO-aware request queuing policy that prioritizes requests to
functions that have a higher likelihood of meeting SLOs, effectively
improving the SLO attainment.

We have implemented and evaluated \SysName through a pilot deployment in \Cloud~\footnote{We have open-sourced \SysName's single-node prototype at \url{https://github.com/FCSLab/torpor}.}, one of the world's largest commercial serverless platforms.
Evaluation results show that \SysName achieves low-latency model inference, comparable with native executions.
\SysName can share a single GPU across hundreds of inference functions and load-balance GPUs with model swapping, resulting in over 10$\times$ cost reduction compared with current GPU offering in \Cloud.
With its efficient SLO-aware scheduling and queuing policies, \SysName can serve 480 functions on a 4-GPU worker node while achieving low tail latency and satisfying millisecond-scale SLOs for all functions.
Cluster experiments further demonstrate that \SysName scales well with the number of inference functions at low resource cost and meets per-function latency SLOs for thousands of functions.
\SysName has been beta-released in a pilot production cluster in \Cloud, saving
70\% of user costs on average and 65\% of GPU provisioning costs for \Cloud.

\section{Background and Motivation}
\label{sec:background}

\revise{
In this section, we first give an overview of serverless inference.  We then describe the inefficiency of existing solutions to enabling GPUs in serverless platforms, and highlight four key requirements
in this regard.
}



\subsection{Serverless Inference}
\label{sec:background-serverless}

\begin{figure}
    \centering
    \includegraphics[width=0.42\textwidth]{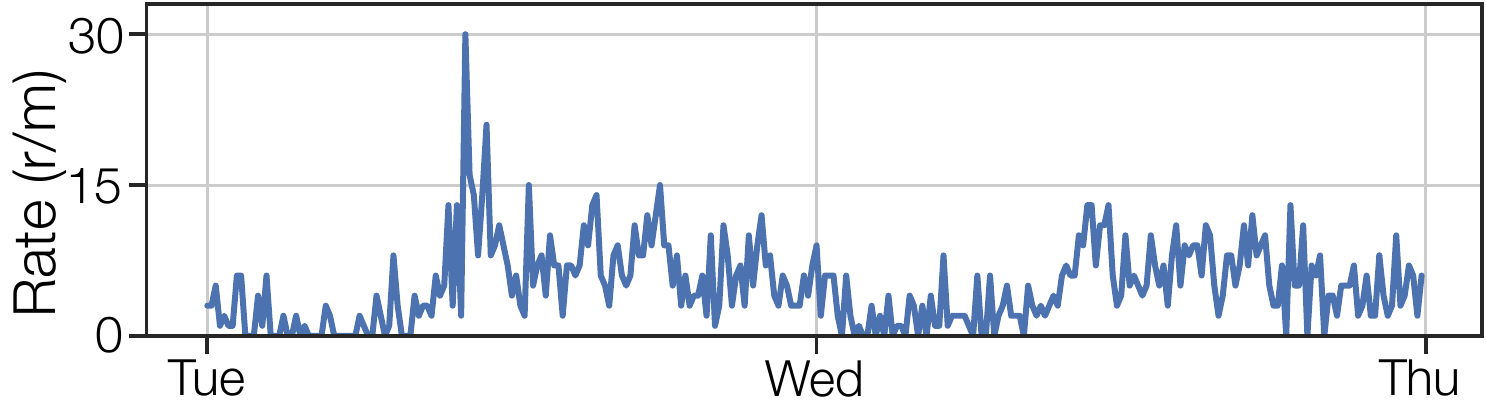}
    \caption{A two-day request trace of a typical GPU inference function in \Cloud.}
    \label{fig:gpu_inf_example}
	\vspace{-.2in}
\end{figure}

As a leading serverless platform with a global presence, our \Cloud has observed
a growing adoption among enterprise customers who opt to deploy their inference services using serverless functions, known as \emph{serverless inference}.  
In comparison to existing inference services based on a ``serverful'' cloud model, such as AWS SageMaker~\cite{sagemaker}, serverless inference significantly alleviates the burden of server management for cloud users.
Specifically, the serverful approach requires users to manually configure various system-level parameters (e.g., VM types, GPUs, CPU cores, etc.) and manage resource provisioning (e.g., scaling the number of VMs up or down according to demand changes).
In contrast, serverless inference enables users to simply publish models with inference code as functions, and then cloud providers automatically handle resource provisioning, autoscaling, scheduling, and fault tolerance. 
Furthermore, compared with the serverful approach,  serverless inference also offers substantial cost savings as users do not pay for idle resources under
the pay-per-use pricing model~\cite
{yang_infless_2022,yu_gillis_icdcs,zhang_mark:_2019,ali_batch_nodate}.
In \Cloud, the requests to a function typically exhibit dynamic, bursty arrival
patterns as shown in Fig.~\ref{fig:gpu_inf_example},
consistent with previous research findings~\cite
{shen_nexus_2019,zhang_shepherd_nodate,gujarati_serving_2020,han_microsecond-scale_2022,lee_pretzel,kosaian_parity_2019,romero_infaas_nodate,
crankshaw_clipper,choi_serving_2022}. By leveraging the high elasticity of
a serverless platform, inference functions can quickly scale in response to
the changing workload, while users are billed based on the actual function runtime at a fine granularity, such as 1~ms~\cite
{aws_lambda,azurefunc}.


\subsection{GPU Support in Serverless Platforms}
\label{sec:background-limitations}

\revise{
Despite the benefits of the serverless inference model, existing serverless platforms, including \Cloud and other leading platforms, currently lack efficient support for GPUs, which impedes their ability to achieve high-performance serverless inference.
\Cloud users also have expressed a compelling need to execute their models in GPU-enabled functions.}

\PHM{\revise{Existing solutions and their inefficiency.}}
\revise{A number of recent systems have been proposed to support GPUs in serverless platforms~\cite{ali_fc,heter_serverless_2022,dgsf_2022,yang_infless_2022}.
They, however, still follow the approach of existing serverful model serving systems (e.g.,  Nexus~\cite{shen_nexus_2019} and INFaaS~\cite{romero_infaas_nodate}), and deploy inference models as long-running containers where each container, when created, is bound to a specific GPU (i.e., early binding).
The deployed model remains in the memory of a designated GPU to handle future requests, and the occupied GPU resources can only be reclaimed after the model serving terminates.}

\begin{figure}[tbp]
	\centering
	\includegraphics[width=0.45\textwidth]{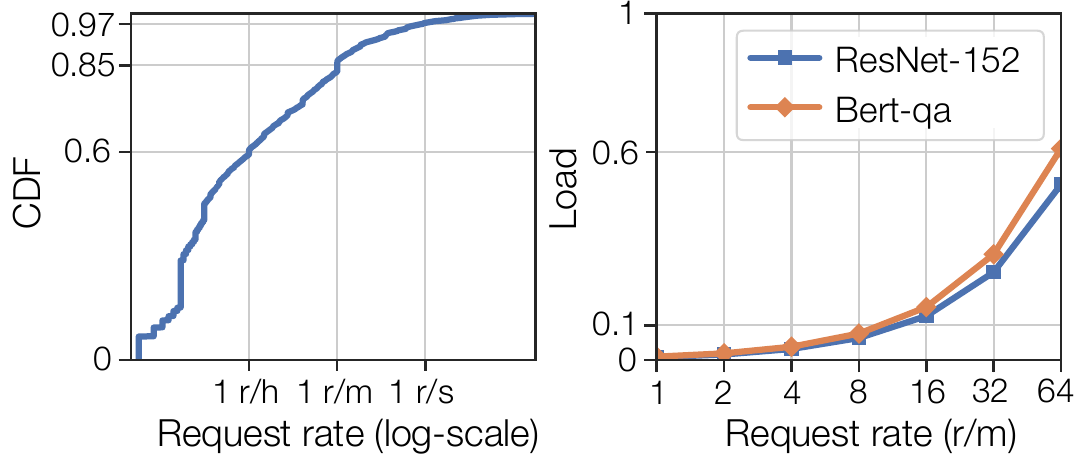}
	\caption{CDF of average function request rates from a one-week production trace (left) and the GPU load under various per-function request rates when running multiple functions on a \texttt{V100} GPU to saturate its 32~GB memory (right).}
	\label{fig:motiv_low_rate}
	\vspace{-.1in}
\end{figure}

\revise{However, the early-binding approach deviates from the serverless paradigm and is costly for both cloud
users and providers. First, binding inference functions to GPUs occupies resources for extended duration, even when idling. 
Thus, users are obligated to pay for the allocated GPUs regardless of actual usage~\cite{fc_billing}, leading to high
expenses that undermine the cost-saving benefits of serverless
inference. Second, this approach results in severe GPU underutilization,
considering the low average request rates of most inference functions and the cross-GPU load imbalancing. Fig.~\ref{fig:motiv_low_rate} (left) depicts the distribution of the
average request rates of \Cloud functions in a one-week trace, revealing that
85\% (97\%) of functions were invoked only once per minute (second) on
average\footnote{For confidentiality reasons, we depict the request rates
of both CPU and GPU functions, which exhibit similar patterns (see Fig.~\ref{fig:gpu_inf_example}).}. These findings align with
the observations from other production traces~\cite
{azurefunc, shahrad_serverless_2020}.
Fig.~\ref{fig:motiv_low_rate} (right) further illustrates that consolidating multiple models to fill GPU memory can still lead to low GPU load.
Meanwhile, packing models into a GPU can cause temporary overloading due to the bursty request patterns (Fig.~\ref{fig:gpu_inf_example}), thus inevitably leading to hotspots and load imbalancing in a multi-GPU setting.
The impact of load imbalancing will be shown in Fig.~\ref{fig:swap_cross_comp} in \S\ref{sec:eval_model_swap}.
}

\begin{table}[t]
    \setlength{\tabcolsep}{3pt}
    \centering
    \caption{\shep{Model startup times (s) under \SysName (runtime isolation mode in \S\ref{sec:system_isolation}) and cold-starts. \SysName's startup time is broken down into model loading and runtime resumption.}}
    \footnotesize
    \label{tab:runtime_isolation}
    \begin{tabular}{ccccc}
        \toprule
        \multirow{2}{*}{\textbf{Model}} & \multicolumn{2}{c}{\textbf{\SysName}} &  \multirow{2}{*}{\textbf{Cold-start}} & \multirow{2}{*}{\textbf{Mem. footprint}} \\
        \cmidrule(lr){2-3} 
        & \textbf{Model}  & \textbf{Runtime} & \\
        \midrule
        ResNet-152~\cite{resnet}  & 0.03 & 0.26 & 8 & 1.6~GB \\
        Bert-qa~\cite{devlin2019bert} & 0.14  & 0.19 & 11 & 2.4~GB  \\  
        Stable Diffusion~\cite{sdiff} & 0.24 & 1.5 & 25& 5.1~GB\\
        Llama3-8B~\cite{llama3} & 1.6 & 1.4 & 48 & 13~GB\\
        Qwen-14B~\cite{qwen} & 2.1 & 1.5 & 57 & 20.1~GB\\
        Llama2-13B~\cite{llama2} & 2.5 & 1.9 & 61 & 24.5~GB\\
        \bottomrule
    \end{tabular}
	\vspace{-.2in}
\end{table}

\add{
To reduce costs, current systems need to frequently reclaim GPU resources when functions are inactive, avoiding charges for unused GPUs and allowing other functions to utilize idle resources. 
Unfortunately, this approach leads to frequent function cold starts, leading to significant overhead for model inference.
Table~\ref{tab:runtime_isolation} shows model startup times under cold starts, which need tens of seconds for GPU container setup, ML framework startup, GPU runtime creation, and model initialization\footnote{We exclude the delay of fetching a remote container image or a model file for cold starts, which can take extra seconds to minutes to complete~\cite{wang_faasnet_nodate}. A detailed discussion of \SysName's performance is provided in \S\ref{sec:pilot}.}.
Therefore, the cold-start overhead far exceeds the typical SLO requirement of model inference. 
}

\setlength{\tabcolsep}{2.9pt}
\begin{table}[t]
    \centering
    \caption{\revise{A comparison of \SysName and existing solutions that offer GPU support on serverless platforms.}}
    \footnotesize
    \label{tab:comp_property}
    \begin{tabular}{ccccc}
        \toprule
        \textbf{Solution} & \makecell{\textbf{GPU} \\ \textbf{pay-per-use}} & \makecell{\textbf{GPU} \\ \textbf{efficient}}  & \makecell{\textbf{SLO} \\ \textbf{compliant} }& \makecell{\textbf{Model}\\\textbf{agnostic}} \\
        \midrule
        Alibaba Cloud\cite{ali_fc} & $\times$  & $\times$  & $\times$ &  $\checkmark$ \\  
        Molecule\cite{heter_serverless_2022}  & $\times$  & $\times$ & $\times$ &  $\checkmark$ \\
        DGSF\cite{dgsf_2022}  &$\times$  & $\times$ &  $\times$ &  $\checkmark$ \\
        INFless\cite{yang_infless_2022}   &  $\times$  &$\times$ & $*$ &  $\times$ \\
        \textbf{\SysName} & $\checkmark$ &  $\checkmark$  & $\checkmark$ &  $\checkmark$ \\
        \bottomrule
    \end{tabular}
    \vspace{-.2in}
\end{table}

\PHM{Requirements of serverless inference.}
\add{
Table~\ref{tab:comp_property} summarizes key requirements of serverless inference and compares \SysName with other existing solutions.
Serverless users should be billed only when their functions are invoked and running on GPUs to achieve
substantial cost savings (\emph{\textbf{pay-per-GPU-use}})\footnote{In our experiences, enterprise customers are willing to pay a nominal fee to
retain idle functions in host memory for substantially improved performance (\S\ref{sec:pilot}), similar to the function keep-alive charge meant to avoid cold starts~\cite{aws_provisioned,jia_nightcore_2021,fc_billing}.}.
Serverless platforms like \Cloud should serve as many inference functions as possible using a minimum number of GPUs, thereby attaining high GPU utilization (\emph{\textbf{GPU efficient}}). 
The platform should allow users to specify their latency
SLOs and strive to meet the latency SLOs for all functions (\emph{\textbf{SLO compliant}}).
For confidentiality reasons, the serverless platform should avoid inspecting detailed model structure, which can be of high business value (\emph{\textbf{Model agnostic}}).
}

Compared with \SysName, none of existing solutions can meet all desired requirements. \Cloud and Alibaba Function Compute~\cite{ali_fc} are leading commercial
 serverless platforms with GPU supports; Molecule~\cite
 {heter_serverless_2022} introduces a serverless platform that supports GPUs
 and other hardware devices; DGSF~\cite{dgsf_2022} enables serverless
 functions to access GPUs in a remote cluster. 
These systems employ the early-binding approach as previously discussed, failing to enable pay-per-GPU-use billing and achieve high GPU efficiency.
Moreover, they are oblivious to the semantics of model inference and unable to meet latency SLOs. 
INFless~\cite{yang_infless_2022} presents a serverless inference system that early-binds functions to GPUs.
While INFless proposes function scheduling and keep-alive schemes aimed at low-latency inference, it still leads to function cold starts and SLO violations (details in \S\ref{sec:eval_node}).
Furthermore, INFless requires model knowledge for operator-level profiling.
We leave more discussions on related work to \S\ref{sec:discuss}.

\section{Key Insight and Challenges}
\label{sec:overview}

\PHM{\shep{Key insight.}}
\shep{
As described in \S\ref{sec:background-limitations}, the current early-binding approach of retaining inference models in GPU memory leads to high idling costs and underutilized resources.
Therefore, an efficient serverless inference platform should enable \emph{late binding}, where GPUs are managed as a resource pool and idle inference models reside in host memory, dynamically swapping into any available GPUs upon request.
This approach should also be \emph{easily deployable} on real-world serverless platforms without requiring intrusive changes.
Late binding offers several key advantages in meeting the requirements in \S\ref{sec:background-limitations}.
\textbf{First}, keeping models in host memory eliminates GPU memory usage during idle periods, enabling pay-per-GPU-use billing and cost savings for cloud users.
\textbf{Second}, host memory is significantly larger than GPU memory (e.g., a few TB vs. tens of GB), allowing for consolidation of multiple low-frequency functions onto a single GPU with improved GPU utilization.
Late binding also facilitates load-balancing across multiple GPUs in a pool.
\textbf{Third}, model swapping provides an efficient method to resume function execution compared to cold starts in the early-binding approach, thereby facilitating SLO compliance.
\textbf{Finally}, late binding can be performed transparently to users within the GPU pool, which holds a holistic view of memory usage without requiring detailed model-specific knowledge.
}

\PHM{\revise{Challenges.}}
\add{
Implementing GPU pooling and late binding in the serverless platform presents three challenges.
\emph{\textbf{C1: Efficient GPU pooling and model swapping}}. GPU pooling requires inference functions to synchronize with a remote GPU pool~\cite{gVirtuS_paper,rcuda}, which introduces additional communication overhead compared to local executions and presents challenges in achieving low-latency inference.
\emph{\textbf{C2: GPU memory management.}} 
To enable seamless late binding, the platform should automatically monitor and manage memory usage without detailed model knowledge.
This requires a unified and efficient GPU memory management system across the GPU pool.
\emph{\textbf{C3: SLO compliance and resource efficiency}}. The platform should provide efficient request scheduling and model placement algorithms that effectively utilize the late binding mechanism to meet latency SLOs and enhance resource efficiency.
}

In the following sections, we present \SysName, a GPU-enabled serverless platform that addresses the aforementioned challenges and, importantly, is readily-deployable onto real-world serverless platforms without intrusive changes.
\section{\SysName System Design}
\label{sec:system}


\subsection{Architecture overview}
\label{sec:system_arch}


\begin{figure}
    \centering
    \includegraphics[width=0.32\textwidth]{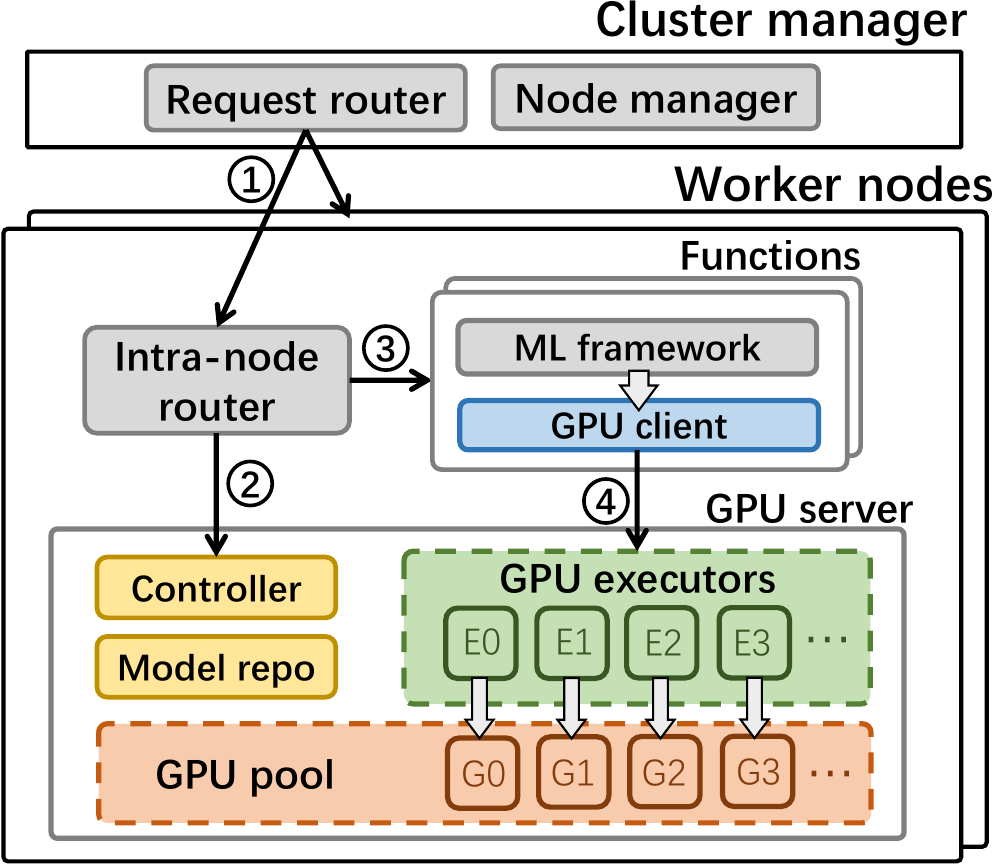}
    \caption{Architecture overview of \SysName. A request arriving at a \SysName cluster is first routed to a worker node hosting its target function \circledNum{1}. The router in the worker node synchronizes with the GPU server to query the executor for this request \circledNum{2}, and then routes it to the function instance with the target executor ID \circledNum{3}. The function instance next processes the request and uses a GPU client to automatically redirect CUDA API calls to this executor \circledNum{4}, and finally returns the result to the user after request completion.}
    \label{fig:system_arch}
	\vspace{-.2in}
\end{figure}

\shep{
Fig.~\ref{fig:system_arch} provides an overview of the architecture of \SysName, which comprises two main components: the cluster manager and worker nodes.
The cluster manager handles cluster-level tasks, including request routing, node allocation, and resource scaling. 
It dynamically schedules function instances and routes inference requests to maintain load balancing across worker nodes and ensure fault tolerance (\S~\ref{sec:system_isolation}). 
At each worker node, \SysName employs GPU pooling, where a GPU server manages all local GPUs as a pool and allows functions to dynamically access any available GPUs.
Within the GPU server, a model repository manages models in host memory; GPU executors handle CUDA execution, perform necessary model swapping, and manage GPU memory on the associated GPUs;
the controller maintains a global view of GPU memory and executor status, and decides how to schedule requests to executors. Additionally, each worker node runs an intra-node router to signal the GPU server about request arrivals and route requests to local inference functions.
Once the target function receives a request, it interacts with the scheduled executor through a GPU client by remoting CUDA API calls.
All components within the worker node---the GPU server, intra-node router, and functions---are deployed as containers. 
}

Key to \SysName is to develop an efficient GPU server that enables low-latency model inference and addresses the challenges discussed in \S\ref{sec:overview}.
Specifically, we will elaborate upon \SysName's designs to address the following questions:
1) how \SysName achieves low-latency GPU pooling (\S\ref{sec:system_remote}) and model swapping (\S\ref{sec:system_swap});
2) how \SysName tracks the memory footprint of functions and manages GPU memory (\S\ref{sec:system_memory});
3) how \SysName ensures isolation and handles failures (\S\ref{sec:system_isolation}).

\subsection{GPU Remoting}
\label{sec:system_remote}

\PHM{Asynchronous API redirection.}
Existing GPU remoting solutions~\cite{rcuda,cu2rcu} introduce high
communication overhead due to synchronizations for individual API calls
during model inference (details in \S\ref{sec:eval_remoting}).
Leveraging the computing pattern of model inference, \SysName proposes \emph{asynchronous} API redirection to reduce synchronizations.
Specifically, we observe that the intermediate steps in an inference execution are typically performed asynchronously on the GPU, where intermediate data is generated and consumed in GPU memory without requiring data transfer to the host until the execution completes.
Consequently, a function can redirect intermediate CUDA calls to the GPU executor asynchronously without waiting for their results, and only perform synchronizations for the final output.
This approach preserves the execution order of CUDA APIs, ensuring the correctness of 
the inference results.

Following this insight, \SysName categorizes CUDA APIs into two groups based on their semantics: 1) \emph{synchronous, blocking} APIs that require the host to await their completion before proceeding, such as \texttt{cudaMalloc}; and 2) \emph{asynchronous, non-blocking} APIs that do not alter the host's runtime state, such as \texttt{cudaLaunchKernel}, which allows for asynchronous API redirection. 
Most APIs issued during inference fall into the asynchronous category, presenting opportunities to mitigate the synchronization overhead.
\SysName can further batch consecutive CUDA API calls to enhance asynchronous API redirection 
(see APIs and batching details in Appendix~\ref{sec:appendix_cuda_api}).

\subsection{Model Swapping}
\label{sec:system_swap}


As serverless platforms are constrained from examining the detailed model structures, it poses challenges to achieve seamless and efficient model swapping.
\SysName overcomes the problem by leveraging two insights: 1) tracking general memory footprint of model inferences is feasible within a GPU pool, and 2) the memory access pattern of a model---the addresses and access order of model parameters---generally remains consistent across requests.
Therefore, \SysName only tracks the first function execution (i.e., cold start), and applies the pattern to future request executions (see memory tracking in \S\ref{sec:system_memory}).
\SysName performs model swapping on demand at the request level, and enhances performance through pinned memory pool and pipeline execution.


\begin{figure}
    \centering
    \includegraphics[width=0.33\textwidth]{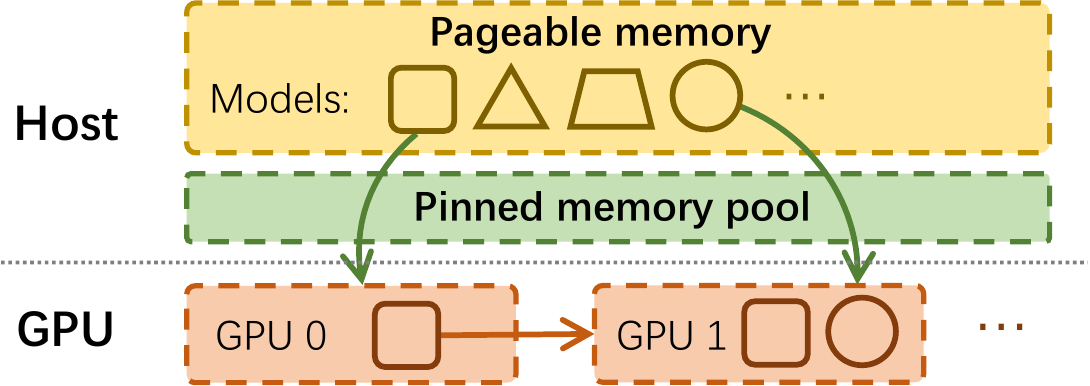}
    \caption{An example of model swapping. Models can be swapped from host to GPU through PCIe (green arrows), or across GPUs through NVLink (red arrow).}
    \label{fig:system_swap}
	\vspace{-.2in}
\end{figure}

\PHM{\revise{Model swapping and pipeline execution.}}
\add{
Fig.~\ref{fig:system_swap} illustrates the model swapping in \SysName, where it utilizes pinned memory to enhance the host-to-GPU model loading and supports fast, cross-GPU model swapping via high-speed NVLink.
\SysName's swapping can be generally applied to various types of models, including those that require maintaining runtime states (e.g., KV cache in LLMs) by treating these states as part of the model itself. 
\SysName further employs pipeline execution to enhance swapping performance, which is particularly effective for models computed in one forward pass, allowing the transmission of subsequent layers to overlap with the computation of previous layers.
}

Key to pipeline execution is to judiciously group model parameters for swapping.
Grouping too few parameters triggers a large number of transmissions and high synchronization overhead; conversely, grouping too many parameters impairs pipeline efficiency.
Pipeline strategies in previous systems like PipeSwitch~\cite{bai_pipeswitch_nodate} and DeepPlan~\cite{deepplan_2023} do not suit serverless inference as they assume model structure is provided and require extensive model profiling.
\SysName employs a model-agnostic approach to determine the group size for model pipelining.
We observe that the transmission performance experiences an "elbow point" concerning group sizes: increasing the group size improves overall transmission throughput, but the improvement becomes marginal after a certain point. 
We therefore select this elbow point as the group size, which can achieve good swapping performance without substantially compromising pipeline efficiency.
This group size depends only on hardware configurations such as PCIe, and can be easily determined by profiling various-sized data (e.g, about 2 MB in our testbed).
This approach requires no detailed model knowledge and can be directly applied to various models.

\PHM{Model eviction.}
We observe that model unloading from GPUs to host can result in considerable overhead and can interfere with concurrent inference executions. 
Therefore, \SysName always maintains a copy of the model in the host, and only invalidates its GPU memory region during eviction. 

\subsection{Memory Management}
\label{sec:system_memory}

\SysName presents two key requirements for GPU memory management, as compared with other systems~\cite{mem_management_url,peng_capuchin_2020, choi_memory_nodate}.
First, late binding requires a pool of GPUs to share the same logical memory space.  This is because the inference functions do not recognize backend GPUs, and consistently access models using identical memory addresses even across different GPUs.
Second, model swapping triggers frequently GPU memory (de)allocation, which leads to substantial overhead when using native methods like \texttt{cudaMalloc} (see Fig.~\ref{fig:eval_block_mem}).
We therefore design a GPU memory management system that can effectively hide memory address differences between various GPUs and provide low-latency memory (de)allocations.


\PHM{Memory address management.}
We observe ML frameworks like PyTorch typically organize data into blocks, each containing multiple parameters.
\SysName leverages such memory layout to perform memory mapping at the block level.
In particular, \SysName monitors memory blocks for each function and maintains a mapping to their actual physical addresses after model swapping. 
Since the internal data layout within each block remains unchanged (e.g., parameter offsets), \SysName can easily obtain the physical address of a parameter using its associated block address and offset. 
This approach eliminates the need for extensive metadata maintenance for individual data pointers, enabling the efficient address translation with low management overhead.

\PHM{Memory block allocation.}
To mitigate the high overhead of native GPU memory allocation, \SysName reserves all GPU memory at bootstrap and internally manages memory blocks. 
This provides a shim layer to service memory requests from functions, without needing the native method.
Key to this approach is to avoid memory fragmentation, which can decrease available GPU memory and harm overall efficiency.
\SysName effectively addresses this issue by extending the Buddy memory allocation scheme~\cite{buddy_memory} and leveraging unique characteristics of inference.
It consolidates memory blocks from the same models to minimize fragmentation and enables sharing of common-sized blocks across different models 
(see details in Appendix~\ref{sec:appendix_block_alloc}).

\subsection{Isolation and Fault Handling}
\label{sec:system_isolation}


\PHM{Resource isolation and GPU runtime management.}
\add{
\SysName provides container-level isolation for CPU and memory resources\footnote{\SysName makes no assumption on function sandboxes and can also support microVMs\cite{agache_firecracker_2020,ustiugov_benchmarking_2021}.}, similar to existing serverless platforms~\cite{ali_fc,tian2022Owl,pheromone}.
For GPUs, \SysName executes only one function on a GPU at a time and isolates GPU memory regions across functions. This is achieved through \SysName's GPU server, which has full control over CUDA API execution and GPU memory access.}

\add{
\SysName offers two isolation modes for GPU runtime: 1) runtime sharing, which runs a single runtime on a GPU for multiple models, for instance, in a more trusted environment, and 2) runtime isolation, which maintains a dedicated runtime for each model and suits better for a more untrusted environment.
By default, \SysName employs runtime sharing to improve resource efficiency. 
When stricter isolation is necessary, \SysName can switch to runtime isolation mode.
Indeed, in our pilot deployment (\S\ref{sec:pilot}), \SysName runs in the runtime isolation mode. 
The overhead incurred by runtime isolation is generally acceptable, e.g., hundreds of milliseconds to 1.5 seconds as shown in Table~\ref{tab:runtime_isolation}, which is still over one order of magnitude latency improvement compared with cold starts.
}


\PHM{Fault handling.}
\add{
\SysName sustains various system component failures.
In case of function failures, \SysName restarts them to resume the execution. 
For executor failures or GPU runtime errors, \SysName migrates the affected models to other working GPUs (executors) via swapping, and then restarts the failed ones.
When runtime isolation is employed, \SysName can ensure that buggy function executions do not affect others, achieving stronger fault isolation.
The GPU server also persists runtime states (e.g., models and metadata) in local storage to allow fast recovery from an entire failure of the GPU server.
}

At the cluster level, \SysName persists metadata of individual nodes in a database, which enables the cluster manager to retain these states and recover from failures, aligning with current practices in \Cloud.
It also keeps periodic health checks with the router on each worker node, and handles node failures by launching a new node and migrating all relevant functions.

\section{\SysName Policy Design}
\label{sec:algorithm}

We present how \SysName meets the latency SLOs and delivers resource efficiency (i.e., Challenge \emph{C3} in \S\ref{sec:overview}).
We start with the design overview, followed by individual policies.

\subsection{Design Overview}
\label{sec:algo_overview}

\PHM{Objective.}
The objective of \SysName's policy design is to meet latency SLOs for inference functions while minimizing the resource cost.
We define a function to comply with latency SLOs if its tail request latency is not longer than a user-specified deadline, and meter the resource cost by the number of worker nodes.
Key to achieving this goal is to \emph{maximize the number of SLO-compliant functions} at each worker, such that \SysName can efficiently exploit per-worker GPU resources to host as many functions as possible, which in turn reduces the total number of workers required.



\begin{table}
    \centering
    \caption{Latency (ms) of model pipelining execution when concurrently swapping other models through PCIe. The diagonal values indicate the latencies without concurrent models.}
    \footnotesize
    \label{tab:swap_interference}
    \begin{tabular}{cccc}
        \toprule
        \textbf{Model} & \textbf{DenseNet-169} & \textbf{ResNet-152} & \textbf{Bert-qa} \\
        \midrule
        DenseNet-169  &  \textbf{27} & 27 (+0\%) & 27 (+0\%) \\ 
        ResNet-152  &  31 (+7\%) & \textbf{29} & 43 (+48\%) \\ 
        Bert-qa  &  166 (+11\%) & 240 (+61\%) & \textbf{149} \\ 
        \bottomrule
    \end{tabular}
	\vspace{-.2in}
\end{table}

\PHM{Challenges.}
Previous systems have proposed various schemes to meet latency SLOs~\cite{yang_infless_2022,zhang_mark:_2019,gujarati_serving_2020,zhang_shepherd_nodate}; however, their policies do not apply to \SysName for two reasons.
First, previous systems like INFless~\cite{yang_infless_2022} and Shepherd~\cite{zhang_shepherd_nodate} assume sufficient GPU memory and employ early binding, so they schedule model serving instances to GPUs and then batch and route requests to them. 
In contrast, \SysName focuses on late-binding the often lower-frequency or varying-demand functions to a pool of memory-constrained GPUs, requiring a joint design of model management (i.e., model swapping and eviction) and request scheduling. 
Second, previous systems assume a stable model inference latency ~\cite{gujarati_serving_2020,zhang_shepherd_nodate,zhang_mark:_2019,yang_infless_2022,romero_infaas_nodate} which, however, does not hold in our setting --- model swapping can cause unpredictable performance due to PCIe bandwidth contention~\cite{topo_gpu_2017,jiang2020unified}.
For instance, as show in Table~\ref{tab:swap_interference}, concurrently swapping two models through PCIe increases individual model inference latency compared with running them alone, especially for large models (e.g., Bert-qa).

We propose three policies to address the aforementioned challenges.
First, considering that packing many functions together can cause short-term overloading and request queueing, \SysName introduces a request prioritization policy to maximize the number of SLO-compliant functions (\S\ref{sec:algo_queueing}).
Second, \SysName designs a request scheduling and model swapping policy to reduce bandwidth contention across concurrent models, thereby improving overall inference performance (\S\ref{sec:algo_scheduling}).
Lastly, by leveraging the characteristics of model swapping, \SysName proposes an effective model eviction policy to reduce bandwidth footprint in model swapping; combined with the request scheduling policy, \SysName minimizes the interference among concurrent model executions (\S\ref{sec:algo_eviction}).

\subsection{Request Queueing}
\label{sec:algo_queueing}

To maximize the number of SLO-compliant functions, \SysName needs to monitor the SLO compliance of individual functions to determine their request executions.
Intuitively, \SysName prioritizes functions with a higher probability to comply with SLOs.
However, realizing this approach requires answering two questions: 1) how to quantify the likelihood of SLO compliance for a function, and 2) how to determine function execution order for improved SLO compliance.


\SysName proposes a metric, \emph{required request count} (RRC), to measure the ``degree of needed effort'' to meet SLOs.
RRC represents the expected request number that a function needs to successfully serve in order to satisfy SLOs.
Let $n$ be the current number of requests to a function, and $m$ be the number of requests served within deadlines out of $n$.
The RRC of this function is defined as $\frac{pn - m}{1 - p}$, where $p$ is the tail percentile specified in SLOs such as 98\%. 
This is derived from the equation: $\frac{m + RRC}{n + RRC} = p$.
Smaller RRC values indicate a higher likelihood of SLO attainment.
Hence \SysName divides functions into high- and low-priority groups based on their RRCs, and prioritizes their requests accordingly.
We develop an effective strategy to determine the boundary (i.e., RRC threshold) between the two groups that can dynamically adjust function prioritization based on the current load at a worker, thereby improving the overall SLO compliance 
(details in Appendix~\ref{sec:appendix_auto_config}).

\begin{figure}
    \centering
    \includegraphics[width=0.35\textwidth]{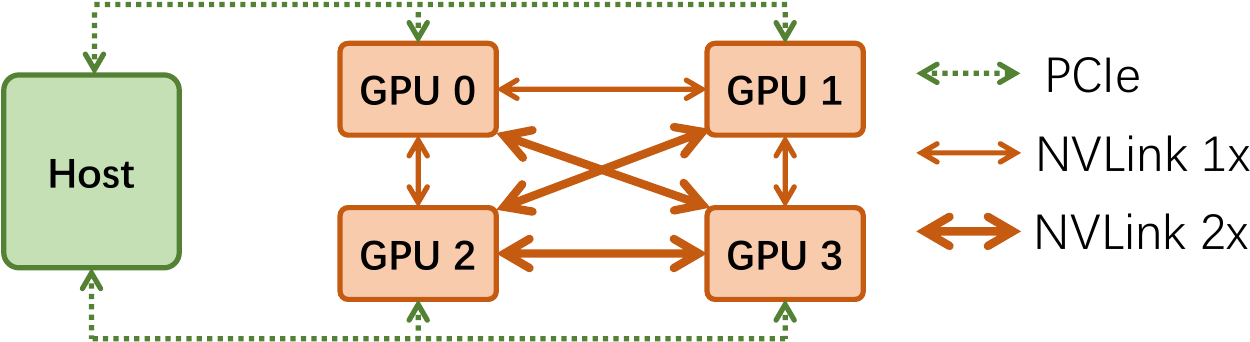}
    \caption{Topology of a 4-GPU worker node in \Cloud.}
    \label{fig:topology}
    \vspace{-.2in}
\end{figure}

\subsection{Scheduling and Model Swapping}
\label{sec:algo_scheduling}


\add{
While model execution latency is often stable, model swapping can incur unpredictable overhead due to PCIe bandwidth contention~\cite{topo_gpu_2017,jiang2020unified}.
Fig.~\ref{fig:topology} shows the topology of a worker node in \Cloud, where each pair of GPUs shares a PCIe switch and GPUs are inter-connected via NVLinks with various bandwidths\footnote{Despite the presence of other GPU interconnects (e.g., NVSwitch in DGX A100), inter-GPU PCIe bandwidth sharing continues to necessitate interference mitigation.}.
The performance slowdown caused by bandwidth contention can vary among models as shown in Table~\ref{tab:swap_interference}, where larger models require more intensive data transmission and exhibit more pronounced performance degradation. 
Hence we propose interfere-aware scheduling to minimize PCIe contention, thereby reducing request latencies.
}

\PHM{Interference-aware scheduling.}
\add{
\SysName exploits the direct NVLink connections between GPUs to reduce PCIe contention whenever possible.
It prioritizes GPU-to-GPU over host-to-GPU model swapping to enable faster model transmission and avoid interference with concurrent PCIe traffic.
When concurrent host-to-GPU swapping is unavoidable, \SysName avoids loading bandwidth-intensive models (e.g., Bert-qa in Table~\ref{tab:swap_interference}) simultaneously to minimize the impact of PCIe contention.
Therefore, models are categorized as heavy or light based on their bandwidth requirements (see Table~\ref{tab:eval_remote_swap}).
}

\begin{algorithm}[t]
	\caption{Interference-Aware Request Scheduling}
    \label{algo:schedule}
    \footnotesize
    \begin{algorithmic}[1]
        \Function{Schedule}{req $r$}
            \State{$A \gets $ set of available GPUs} \Comment{$A \neq \varnothing$, otherwise queueing $r$}
            \State{$M \gets $ set of GPUs hosting the target model}

            \If{$M \neq \varnothing$}
                \State{$G \gets M \cap A$}
                \If{$G \neq \varnothing$} 
                    \State{$g \gets $ any GPU in $G$}
                    \State{\textbf{Execute} $r$ on $g$} \Comment{No swapping}
                \Else
                    \State{$(g, m) \gets $ GPU pair with fastest NVLink, $g \in A, m \in M$}
                    \State{\textbf{Execute} $r$ on $g$; \textbf{Swap} model from $m$} \Comment{GPU-to-GPU}
                \EndIf
            \Else
                \State{$g \gets $ a GPU whose neighbor is not loading models, $g \in A$}
                \If{$g$ not found}
                    \State{$g \gets $ a GPU whose neighbor is loading a light model, $g \in A$}
                \EndIf
                \If{$g$ not found}
                    \State{$g \gets $ any GPU in $A$}
                \EndIf
                \State{\textbf{Execute} $r$ on $g$; \textbf{Swap} model from host} \Comment{Host-to-GPU}
            \EndIf
        \EndFunction
	\end{algorithmic}
\end{algorithm}

Algorithm~\ref{algo:schedule} shows \SysName's scheduling and swapping mechanisms.
\SysName first checks whether the target model is loaded on an available GPU, and if so, directly executes it without swapping (line 8).
If the model is hosted by busy GPUs, \SysName then schedules the request to perform GPU-to-GPU swapping, as the source and target GPUs should have a fast NVLink connection (line 11).
Otherwise, \SysName resorts to the host-to-GPU swapping and prioritizes target GPUs whose neighbors are idle or running light models to reduce PCIe contention (line 18).
Altogether, \SysName minimizes the interference and overhead of model swapping for each request, thus providing low inference latency.

\subsection{Model Eviction Policy}
\label{sec:algo_eviction}

\revise{
    Model eviction plays a critical role in reducing bandwidth contention and enhancing the overall inference performance, in conjunction with \SysName's request scheduling policy.
    Unlike traditional cache eviction strategies which primarily aim to minimize the miss rates, \SysName's model eviction policy considers the performance implications of model swapping for different models to facilitate future model loading.
}

\add{
We notice that swapping light models leads to negligible overhead for end-to-end performance compared with heavy ones (Table~\ref{tab:swap_interference} and Table~\ref{tab:eval_remote_swap}).
Therefore, we tend to evict models that have little or no impact on performance when swapping. 
We employ two mechanisms following this insight.
First, \SysName manages memory of all GPUs as a pool to globally optimize model placement, which ensures that each model can have up to one replica among GPUs when GPU memory is full.
This allows for more efficient model caching and reduces host-to-GPU data transmission.
Second, \SysName prioritizes light models in eviction, as swapping them leads to negligible or no PCIe bandwidth contention.
When only heavy models remain, \SysName adopts Least-Recently-Used (LRU) policy to determine their eviction order.
}

\section{Implementation}
\label{sec:implementation}

\add{
We have implemented \SysName for \Cloud, one of the world's leading commercial serverless platforms.
\SysName's GPU server and GPU client were implemented in 4k and 1.5k lines of C++ code, respectively.
Intra-node router and cluster manager were implemented atop the relevant components in \Cloud.
\SysName's late-binding mechanism imposes no intrusive changes to standard cluster management logic, enabling the reuse of \Cloud's existing manager with minimal changes. 
In fact, \SysName has been successfully deployed in a real-world production environment of \Cloud (\S\ref{sec:pilot}).
We provide a container image as a function template based on PyTorch, where the original CUDA libraries (e.g., \texttt{libcudart.so}) are replaced by our GPU clients to enable GPU remoting.
This requires no modification to the PyTorch framework.
}
\section{Evaluation}
\label{sec:eval}

In this section, we evaluate \SysName with the runtime sharing mode using production traces from \Cloud.  We intergrate \SysName with runtime isolation mode into \Cloud's real-world production environment and report the results in \S\ref{sec:pilot}.  

\setlength{\tabcolsep}{2.9pt}
\begin{table}[t]
    \centering
    \caption{\new{Various models and their latencies (ms) with GPU remoting and model swapping. \underline{Underlined} are heavy models where swapping via PCIe slows down the inference (see \S\ref{sec:algo_scheduling}).}}
    \label{tab:eval_remote_swap}
    \footnotesize
    \begin{tabular}{ccccccc}
        \toprule
        \textbf{Model} & \textbf{Native} & \textbf{GPU remoting} & \textbf{Swap-PCIe} &  \textbf{Swap-NVLink}\\
        \midrule
        DenseNet-169 & 30 & 25 & 27 & 26\\
        DenseNet-201 & 36 & 28 & 30 & 30\\
        Inception-v3 & 19 & 14 & 17 & 16\\
        EfficientNet & 17 & 12 & 13 & 13\\
        \underline{ResNet-50} & 11 & 9 & 13 & 11\\
        \underline{ResNet-101} & 20 & 14 & 22 & 16\\
        \underline{ResNet-152} & 25 & 17 & 25 & 20\\
        \underline{Bert-qa} & 42 & 43 & 144 & 45\\

        \bottomrule
    \end{tabular}

	\vspace{-.1in}
\end{table}


\PHM{\shep{Settings.}}
\shep{We deploy \SysName at \Cloud following the realistic production specification of its serverless platform.
\SysName runs in a cluster with up to 6 workers. 
Each worker node has 48 vCPU cores, 384 GB memory, and 4 NVIDIA V100 GPUs each with 32 GB memory.
We use 8 popular ML models for evaluation, as shown in Table~\ref{tab:eval_remote_swap}, and distribute them across inference functions in a round-robin manner.
All functions are warmed up before running the test workloads.
We compare \SysName against Native execution---the default approach in \Cloud---and INFless~\cite{yang_infless_2022}, a state-of-the-art serverless inference system.
}

\PHM{Metrics.}
We focus on the ratio of functions meeting SLOs and the GPU load in the evaluation.
A function complies SLOs if its tail request latency is within a deadline.
By default, we use $98^{th}$ tail latency, and set deadlines for CV models and Bert-qa to 80 ms and 200 ms, respectively.
The GPU load is measured by the proportion of time during which the GPU is processing inference requests.

\subsection{Overhead of \SysName's Late Binding}
\label{sec:eval_remoting}

\add{
Table~\ref{tab:eval_remote_swap} compares the performance of 8 popular models under Native execution and \SysName with its GPU remoting (\S\ref{sec:system_remote}) and model swapping (\S\ref{sec:system_swap}).
For GPU remoting, \SysName adopts efficient, asynchronous API redirection, leading to comparable performance to Native, or even better for CV models. This is because serving these models requires configuring many cuDNN descriptors where the relevant CUDA APIs are executed on the CPU side and do not require GPU resources; thus, redirecting these APIs effectively distributes CPU-side workloads across functions and the GPU server, enabling functions to access
more CPUs and perform parallel computation.
Note that, CPU resources are not the bottleneck in this scenario.
According to our measurements, the CPU utilization of \SysName (native execution) for ResNet-152 and Bert-qa remains at 27.4\% (8.8\%) and 17.2\% (9.2\%), respectively.
For model swapping, \SysName supports efficient pipeline execution through PCIe, and leveraing NVLink further improves performance.
}

\begin{figure}
    \centering
    \begin{minipage}{0.234\textwidth}
        \centering
        \includegraphics[width=\linewidth]{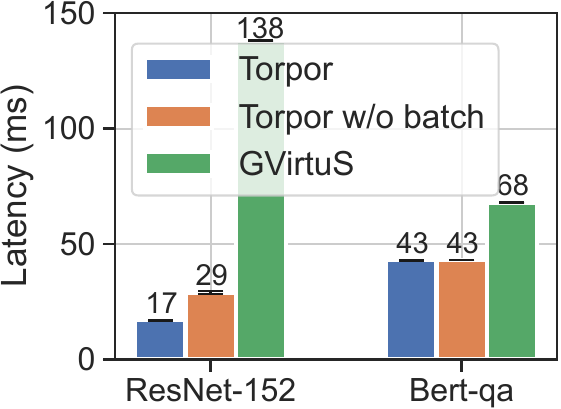} 
        \caption{Inference latency with \SysName and other GPU remoting techniques.}
        \label{fig:eval_gpu_remote}
    \end{minipage}\hfill
    \begin{minipage}{0.234\textwidth}
        \centering
        \includegraphics[width=\linewidth]{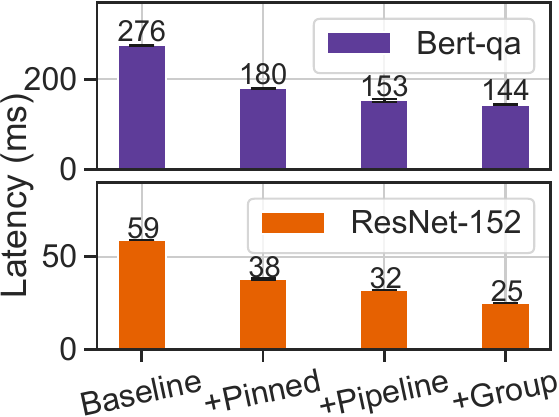} 
        \caption{Performance breakdown of \SysName's model swapping via PCIe.}
        \label{fig:eval_swap}
    \end{minipage}
    \vspace{-.2in}
\end{figure}

\PHM{Performance of \SysName's GPU remoting.}
\new{
To show the advantage of \SysName's GPU remoting, we compare it with GVirtuS~\cite{gVirtuS_paper,gVirtuS_git}, a leading solution among publicly available GPU remoting techniques. 
GVirtuS adopts a synchronous approach to API redirection.
We also evaluate a variant of \SysName that disables call batching in asynchronous API redirection, i.e, ``\SysName w/o batch''.
Fig.~\ref{fig:eval_gpu_remote} shows the inference performance under various approaches. \SysName significantly outperforms GVirtuS and reduces latencies by 88\% and 37\% for ResNet-152 and Bert-qa, respectively.
ResNet-152 triggers a large number of API calls during each inference, leading to high synchronization overhead for GVirtuS.
Asynchronous API redirection (\SysName w/o batch) dramatically reduces the latency by 79\%; with API call batching, \SysName further reduces the latency by 41\%.
Compared with ResNet-152, Bert-qa requires less communication in GPU remoting; therefore, the improvement from asynchronous API redirection is less  but still quite significant.}

\PHM{Performance breakdown of model swapping.}
\new{
To illustrate how each of \SysName's model swapping designs contributes to performance improvement, we break down the inference performance of ResNet-152 and Bert-qa, as shown in Fig.~\ref{fig:eval_swap}.
Specifically, ``Baseline'' directly performs model swapping and then executes inference; ``Pinned'' uses a pinned memory pool for improved swapping performance; ``Pipeline'' overlaps model swapping and execution at the granularity of individual model parameters; ``Group'' groups parameters for efficient pipeline execution. 
We note that enabling pinned memory reduces overall latencies by around 35\%; parameter-level pipeline execution further reduces the latency by 15\%.
By grouping model parameters, \SysName achieves up to 22\% performance improvement over ``Pipeline'', especially for ResNet-152 that consists of many small-sized parameters.}

\subsection{Benefits of \SysName's Late Binding}
\label{sec:eval_model_swap}

\begin{figure}
    \centering
    \includegraphics[width=0.47\textwidth]{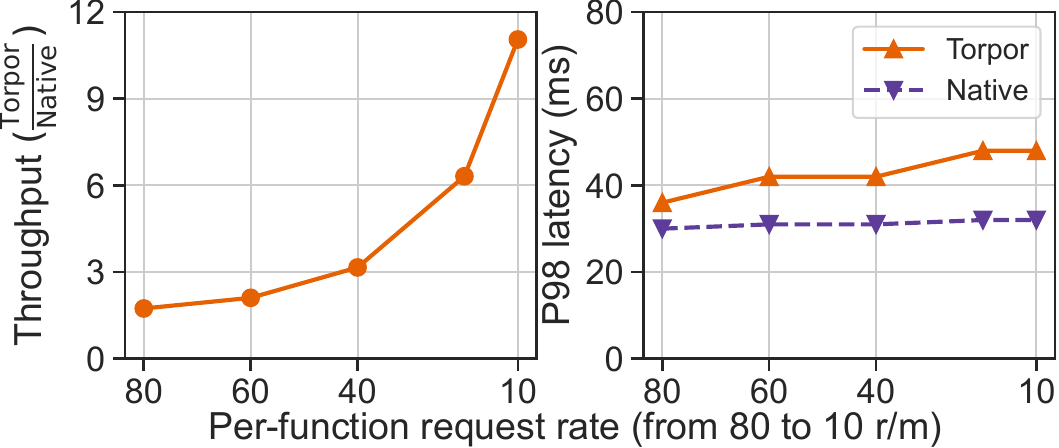}
    \caption{\new{Performance of executing multiple ResNet-152 functions on a single GPU with \SysName's late binding (\SysName) and Native execution.
    We show the throughput of \SysName normalized to Native (left) and the tail latencies under varying per-function request rates (right).}}
    \label{fig:swap_inc_rate}
	\vspace{-.1in}
\end{figure}


\PHM{\shep{GPU efficiency for low-frequency functions.}}
\shep{
With late binding, \SysName substantially reduces per-function memory footprint, thereby enabling the consolidation of many low-frequency functions for improved GPU efficiency.
We stress-test its performance by executing multiple ResNet-152 functions on a single GPU, varying request rates between 80 and 10 requests per minute (r/m)---a typical range of low-frequency functions in production traces (Fig.~\ref{fig:motiv_low_rate}).
In this setup, we execute sufficient concurrent functions on a GPU node to saturate GPU memory and ensure a high overall load.
Fig.~\ref{fig:swap_inc_rate} compares normalized throughput and latency under \SysName and Native executions.
Native's throughput declines as the per-function request rate decreases, due to its limited capacity to host many functions. 
In contrast, \SysName leverages host memory to accommodate many more functions, maintaining high throughput with efficient, request-level GPU sharing. 
For example, \SysName achieves over 10$\times$ higher throughput than Native at 10 r/m.
Furthermore, even when model swapping is required, \SysName still keeps the tail latency below 50 ms.
}

\begin{figure}
    \centering
    \includegraphics[width=0.47\textwidth]{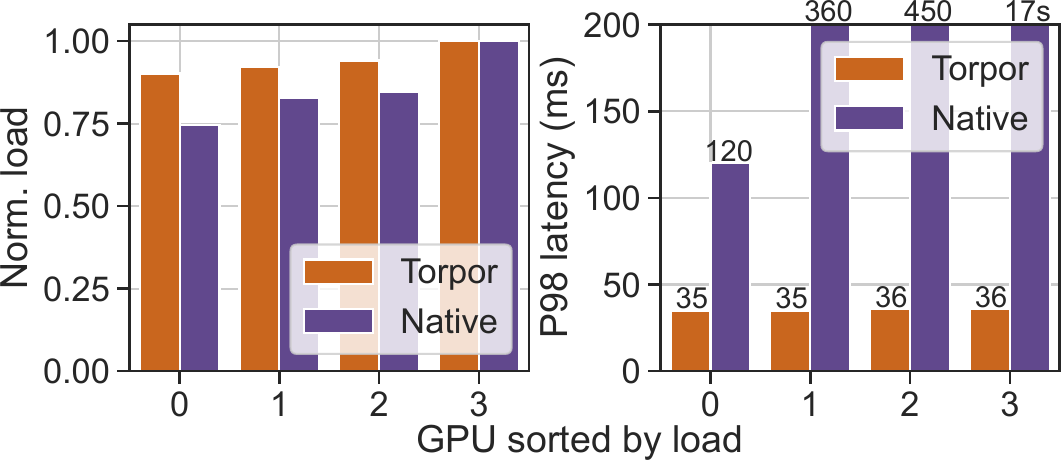}
    \caption{The normalized per-GPU load and the tail request latency with \SysName's late binding (\SysName) and Native.}
    \label{fig:swap_cross_comp}
	\vspace{-.15in}
\end{figure}

\PHM{\new{Cross-GPU load balancing for high-frequency functions.}}
\new{We run 40 high-frequency ResNet-152 functions on a 4-GPU worker, where the average request rate is around 200 r/m.
\new{Fig.~\ref{fig:swap_cross_comp} shows the normalized per-GPU load and the tail request latency with \SysName and Native.
Unlike Native, where GPUs hosting high-frequency functions can easily become overloaded due to bursts of requests, \SysName enables on-demand model migration for efficient load balancing.
Therefore, \SysName achieves much less load variance across GPUs compared with Native, as shown in Fig.~\ref{fig:swap_cross_comp} (left).
Moreover, Fig.~\ref{fig:swap_cross_comp} (right) shows the tail latency of requests executed on each GPU, where Native leads to extremely long tail latency (e.g., multi-seconds) due to high queueing delays.
In contrast, \SysName consistantly delivers fast model inference, achieving a tail latency of around 35 ms on all GPUs.}
}

\subsection{\SysName at A Node}
\label{sec:eval_node}

We next evaluate the performance of \SysName at a node.
We use real-world workloads sampled from production traces (Fig.~\ref{fig:motiv_low_rate}), where function request rates range from 5 to 30 r/m.

\begin{figure}
    \centering
    \includegraphics[width=0.47\textwidth]{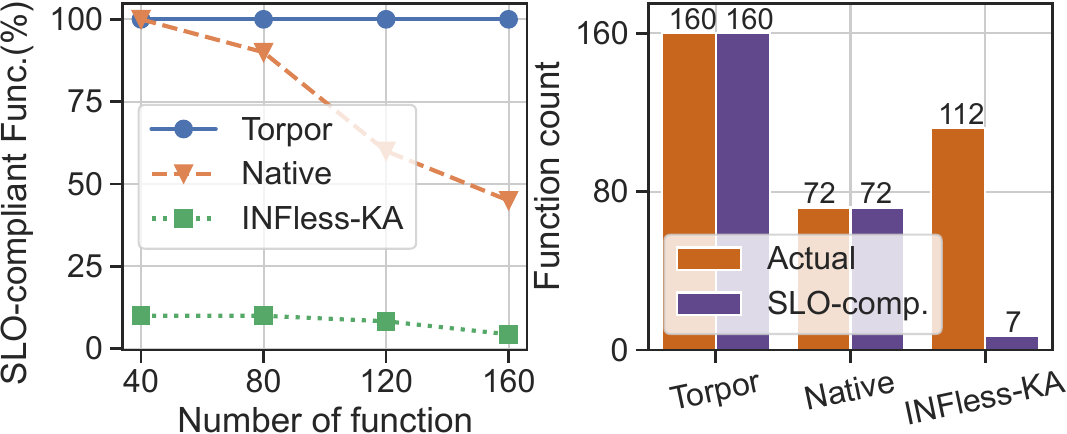}
    \caption{Performance comparison in terms of SLO compliance between \SysName, Native, and INFless-KA.}
    \label{fig:compare_keep_alive}
	\vspace{-.1in}
\end{figure}

\PHM{\revise{Performance comparison.}}
\revise{
We compare \SysName with two baselines, Native execution and INFless~\cite{yang_infless_2022} --- a state-of-the-art serverless inference system. 
INFless introduces a function keep-alive policy to set the lifespan of individual functions based on historical traces, denoted as INFless-KA.
For a fair comparison, we implement the keep-alive policy of INFless (INFless-KA) in the Native system.
Fig.~\ref{fig:compare_keep_alive} (left) shows the ratio of functions meeting SLOs in \SysName, Native, and INFless-KA.  Fig.~\ref{fig:compare_keep_alive} (right) shows the numbers of functions being actually executed and being SLO-compliant, when hosting 160 functions.
With fast model swapping, \SysName executes all 160 functions and also meets SLOs for all functions.
In contrast, due to limited GPU memory, Native can only execute 72 out of 160 functions.  INFless-KA can reclaim GPU memory via cleaning up idle functions, thereby enabling the execution of more functions (i.e., 112 functions) than Native; however, INFless-KA inevitably incurs function cold starts and results in only 7 functions being SLO-compliant. 
}

\PHM{Benefits of \SysName's policies.}
To understand the benefits of \SysName's policy designs, we compare the full \SysName with four baselines.
1) \emph{\SysName-FIFO} uses a FIFO policy in request queueing rather than our SLO-aware policy (\S\ref{sec:algo_queueing}).
2) \emph{\SysName-Random} disables our interference-aware scheduling and model swapping (\S\ref{sec:algo_scheduling}), and randomly schedules a request to an idle GPU if the target model is not loaded, and then triggers model swapping.
3) \emph{\SysName-LRU} adopts a LRU policy in model eviction rather than prioritizing models according to swapping overheads (\S\ref{sec:algo_eviction}).
4) \emph{\SysName-Block} disables our block management policy (\S\ref{sec:system_memory}), and caches the released memory blocks in a single pool;
when a new block is required, it directly returns a cached one in the pool if the requested size can be satisfied, otherwise it frees multiple blocks until the required memory space is available.

\begin{figure}
    \centering
    \includegraphics[width=0.47\textwidth]{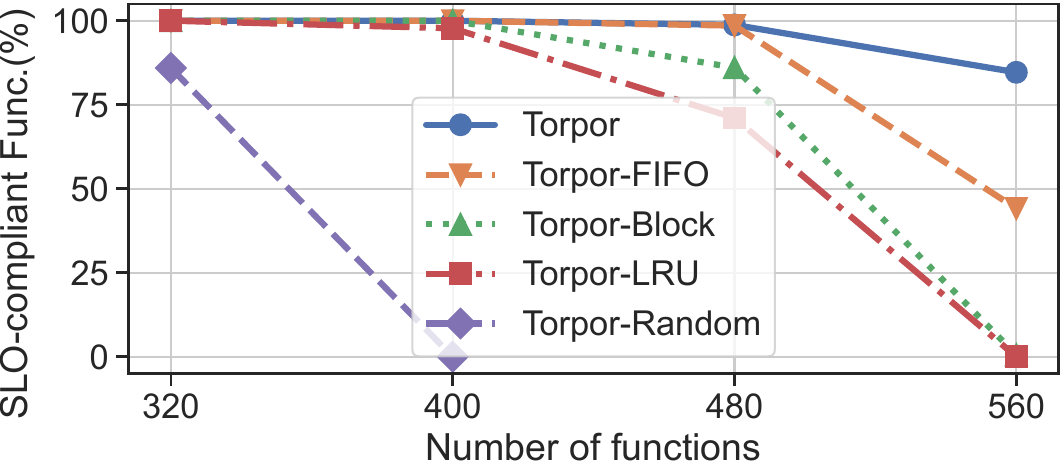}
    \caption{Ratio of SLO-compliant functions with the full \SysName and various different policies.}
    \label{fig:eval_breakdown_overall}
	\vspace{-.15in}
\end{figure}

\begin{figure}
    \centering
    \begin{minipage}{0.235\textwidth}
        \centering
        \includegraphics[width=\linewidth]{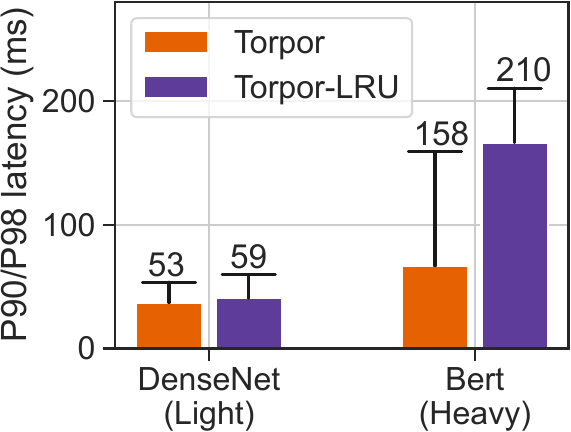} 
        \caption{\add{90$^{th}$ (bars) and 98$^{th}$ (whiskers) tail latencies in \SysName and \SysName-LRU.}}
        \label{fig:eval_light_heavy}
    \end{minipage}
    \begin{minipage}{0.225\textwidth}
        \centering
        \includegraphics[width=\linewidth]{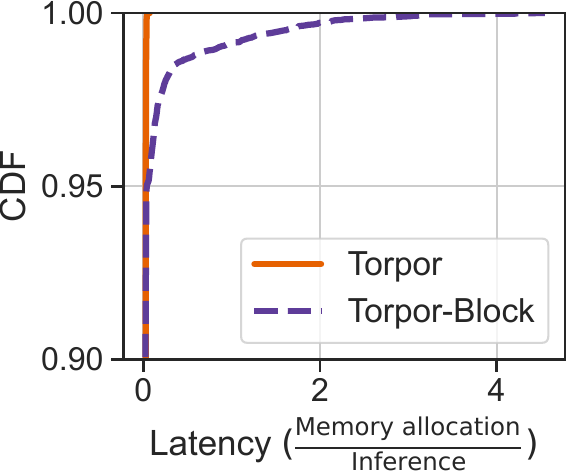} 
        \caption{Latency CDF of memory allocation in \SysName and \SysName-Block.}
        \label{fig:eval_block_mem}
    \end{minipage}\hfill
    \vspace{-.2in}
\end{figure}

Fig.~\ref{fig:eval_breakdown_overall} shows the ratio of SLO-compliant functions.
In particular, \SysName-FIFO is oblivious to SLOs and unable to properly prioritize functions, leading to serious SLO violations when the number of functions is large.  
\SysName-Block cannot reuse various-sized blocks and forces frequent memory allocation via CUDA APIs, which incurs long delay in block allocation and harms overall performance.
\SysName-LRU evicts heavy models often, leading to PCIe bandwidth contention during future model swapping.
\SysName-Random leads to the worst performance due to its inefficient scheduling and model swapping policy, which does not exploit NVLink across GPUs and is oblivious to model heaviness.
Compared with these baselines, \SysName successfully supports over 80\% functions even with 560 functions, maximizing the number of SLO-compliant functions.


\PHM{Efficiency of model heaviness.}
\add{
We evaluate the efficiency of model heaviness with \SysName and \SysName-LRU, using DenseNet-201 and Bert-qa as light and heavy models, respectively.
Fig.~\ref{fig:eval_light_heavy} shows tail latencies of DenseNet-201 and Bert-qa in a mixed workload, where Bert-qa instances account for 60\% of overall memory footprint.
\SysName-LRU is agnostic to model heaviness and can evict heavy models frequently, leading to high tail latencies for Bert-qa that fail to meet its SLOs (200~ms). 
In contrast, \SysName effectively reduces tail latencies of Bert-qa without compromissing performance for DenseNet-201, achieving SLOs for both models.
}


\PHM{Efficiency of memory allocation.}
Fig.~\ref{fig:eval_block_mem} shows the distribution of the latencies of per-request memory allocation normalized to inference time, under \SysName and \SysName-Block.
Due to efficient memory allocation and sharing (\S\ref{sec:system_memory}), \SysName incurs only negligible overhead. 
In contrast, \SysName-Block leads to high allocation overhead (e.g., over 4$\times$ than the actual inference time), harming the end-to-end performance.

\subsection{\SysName in A Cluster}
\label{sec:eval_cluster}

We next evaluate \SysName in a cluster deployment with 6 GPU worker nodes.
As running \SysName in a \Cloud cluster incurs additional system overhead, we set the SLOs for CV models and Bert-qa to 150 ms and 250 ms, respectively.

\begin{figure}
    \centering
    \begin{subfigure}[b]{0.43\textwidth}
        \includegraphics[width=\linewidth]{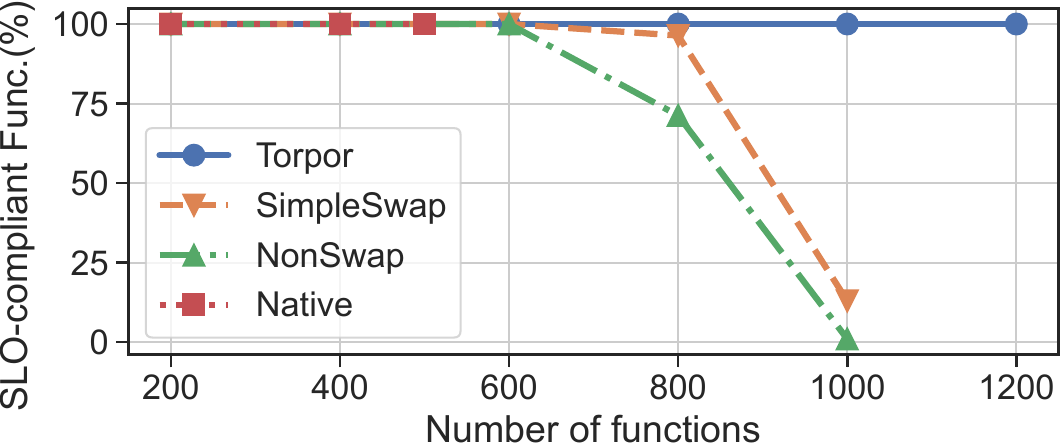}
        \caption{Ratio of SLO-compliant functions under \SysName and baselines.}
        \label{fig:eval_cluster}
    \end{subfigure}
    \hfill
    \begin{subfigure}[b]{0.45\textwidth}
        \includegraphics[width=\linewidth]{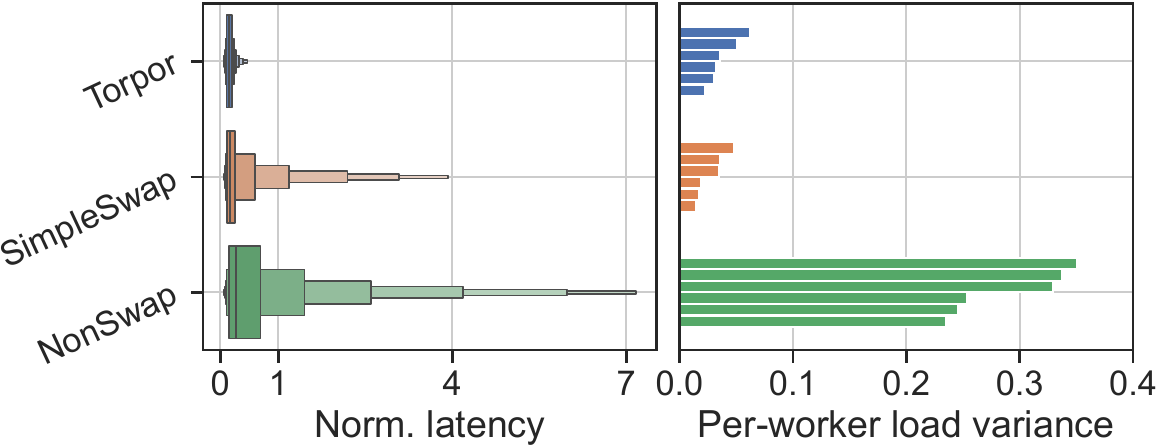}
        \caption{Distribution of per-request latencies normalized to deadlines (left) and the per-worker variance of GPU load normalized to the maximum (right), when running 1k functions. Boxes (left) depict the $1/128$, $1/64$, \dots, $1/2$, \dots, $63/64$, $127/128$ quantiles.}
        \label{fig:eval_cluster_detail}
    \end{subfigure}

    \caption{Cluster evaluation of \SysName.} 
    \label{fig:eval_cluster_all}
    \vspace{-.2in}
\end{figure}

\PHM{Baselines.}
We exclude INFless-KA from our cluster deployment due to its poor performance (see Fig.~\ref{fig:compare_keep_alive}).
We use three baselines:
1) \emph{Native} uses native GPU containers bound to specific GPUs, which is the current practice in \Cloud. 
2) \emph{NonSwap} allows GPU remoting similar to \SysName, but disables model swapping, reducing memory footprint compared with Native.
3) \emph{SimpleSwap} enables model swapping compared with NonSwap.
This approach only supports simple strategies as discussed in \S\ref{sec:eval_node}, including FIFO request queueing, random scheduling, and LRU model eviction.

\PHM{Cluster evaluation.}
Fig.~\ref{fig:eval_cluster_all} compares \SysName with these three baselines.
We first show the ratio of SLO-compliant functions under various number of functions.
As shown in Fig.~\ref{fig:eval_cluster}, only \SysName can consistently meet per-function latency SLOs even with a large number of functions (e.g., over 1000).
Native quickly saturates all GPU memory and only supports up to 500 functions, thus low GPU utilization.
Compared with Native, NonSwap relaxes the constraint of GPU memory and enables more functions; however, it still fixes the binding between functions and GPUs, and causes GPUs overloaded by requests and leads to long tail latency.
Moreover, while SimpleSwap outperforms NonSwap with model swapping, it still suffers from severe SLO violations with a large number of functions (e.g., 1k).

Fig.~\ref{fig:eval_cluster_detail} compares the behaviors of \SysName, SimpleSwap, and NonSwap under 1k functions.
We show the distribution of per-request latencies normalized to corresponding deadlines (left).
In \SysName, almost every request can be served within its deadline, leading to a normalized latency less than 1.
However, SimpleSwap and NonSwap suffer from long tail latency --- over 4$\times$ and 7$\times$ of the respective deadlines.
We also compare the per-worker GPU load of the three systems.
For each worker node, we normalize the loads of its four GPUs to the maximum, and calculate the variance.
Lower variance indicates better load balancing.
Fig.~\ref{fig:eval_cluster_detail} (right) plots the per-worker load variances of three solutions, each with 6 workers in total.
Compared with NonSwap, \SysName and SimpleSwap can effectively balance GPU load across workers with model swapping, thus achieving much less load variance.

\section{\SysName in Pilot Production}
\label{sec:pilot}


\SysName has been deployed in a pilot production cluster
in \Cloud for beta testing. In this section, we present the testing results
and our observations.


\begin{table}
    \centering
    \caption{Overview of \SysName's pilot production.}
    \small
    \begin{tabular}{l|r|l|r}
    \hline
    \textbf{Metric} & \textbf{Value} & \textbf{Metric} & \textbf{Value} \\
    \hline
    \# of users & > 150 & Users' cost savings & 70\% on avg. \\
    \# of GPUs & > 350 & GPU savings & 65\% \\
    \# of daily requests & up to 465k & & \\
    \hline
    \end{tabular}
    \label{tab:production}
    \vspace{-0.1in}
\end{table}

\PHB{Overview of the pilot production.}
\add{
In the deployment, we employ a cost-efficient billing scheme in which users are charged only 10\% of the GPU cost when their functions are inactive and retained in the host memory. 
This billing scheme aligns with \SysName's late binding mechanism and has attracted a variety of real-world inference workloads to \Cloud, including image processing, text generation, and image generation. 
Table~\ref{tab:production} provides an overview of our pilot production system and 
the achieved savings. Currently, \SysName serves over 150 users in a cluster with more than 350 GPUs, handling up to 465k requests everyday.
The system achieves an average cost savings of 70\% for users compared to the previous approach of \Cloud that kept functions long-running on GPUs and billed users for the entire GPU time.
Moreover, \SysName enables \Cloud to consolidate various functions for improved GPU utilization,
resulting in a 65\% reduction in the total number of required GPUs and cost savings for \Cloud.
}

\PHM{\shep{Startup latency.}} 
\shep{Table~\ref{tab:runtime_isolation} shows the performance of \SysName in the pilot production across various models, ranging from CNNs such as ResNet to LLMs like Llama. 
In the production environment, \SysName prioritizes user isolation and manages a dedicated GPU runtime for each function.
Hence we also provide a detailed breakdown of the time required for model loading and runtime resumption in Table~\ref{tab:runtime_isolation}. 
Compared to cold starts in \Cloud, \SysName reduces model startup latencies by over an order of magnitude, e.g., cutting Llama2-13B's startup time to 4.4 seconds---a delay that is acceptable considering that generating all output tokens for a query can take tens of seconds~\cite{fu_serverlessllm_2024,zhong_distserve_2024}.}

\begin{figure}
    \centering
    \includegraphics[width=0.47\textwidth]{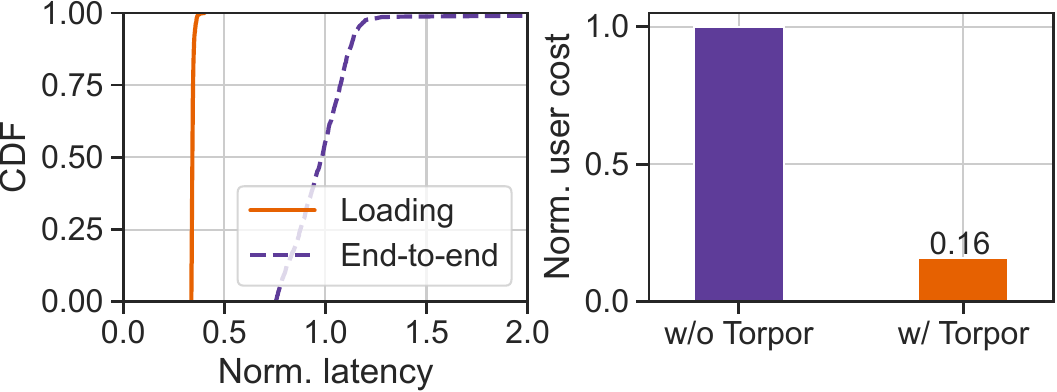}
	\vspace{-.1in}
    \caption{Latency distribution and user cost of a realistic GPU function in \Cloud.}
    \label{fig:prod_case_study}
	\vspace{-.2in}
\end{figure}

\PHM{Case study.} 
We next present a case study of a realistic GPU function in our pilot production, which involves text generation from input images. 
This function is invoked several thousands times per day and follows a request arrival pattern similar to Fig.~\ref{fig:gpu_inf_example}.
Without \SysName, this function would need to be kept long-running on a GPU for low inference latency, resulting in high GPU costs.
Fig.~\ref{fig:prod_case_study} depict latency distribution and user cost of this function.
For confidentiality reasons, we normalize the latency of model (and runtime) loading to the mean of end-to-end times, and the cost to that of long-running GPU functions (i.e., w/o \SysName).
The loading duration is consistent, as shown in Fig.~\ref{fig:prod_case_study} (left), which accounts for only $\sim$30\% of the overall latency.
Fig.~\ref{fig:prod_case_study} (right) illustrates the user cost, where \SysName reduces the total cost by 84\% compared to the previous approach in \Cloud.
Based on this real-world use case, \SysName achieves significant cost savings without compromising end-to-end performance, proving to be an effective solution for serverless inference.

\PHM{\shep{Discussion.}}
\shep{
We have identified three challenges that require further investigation based on our experiences with the production cluster.
\emph{(1) Extremely infrequent functions:} We have observed that some functions are invoked extremely infrequently (e.g., a few requests per hour). However, these functions still result in substantial user costs due to their host memory usage. To address this, we plan to explore the utilization of cheaper storage for low-demand functions and investigate a multi-tier storage architecture that dynamically adapts to request patterns for improved resource and cost efficiency.
\emph{(2) Very large models:} While \SysName's late–binding design is versatile and has been applied to large models (Table~\ref{tab:runtime_isolation}), it currently does not support model-parallel execution across multiple GPUs. 
To accommodate very large models spanning multiple GPUs or nodes, it is crucial to enable efficient model partitioning and parallelization that integrates with \SysName's model-swapping mechanism. 
Therefore, we intend to extend \SysName to leverage various interconnects, including PCIe and NVLink (Fig.~\ref{fig:topology}), for enhanced model parallel and pipeline execution.
\emph{(3) Highly bursty workloads:} While \SysName is primarily designed for efficient resource sharing among many low-frequency functions (Fig.~\ref{fig:motiv_low_rate}), some functions may experience highly bursty workloads with short periods of extensive request arrivals. To handle such spikes, \SysName can proactively maintain host-memory-cached model replicas across multiple nodes, mitigating cold starts at the cost of increased memory usage. 
We plan to explore high-speed inter-node networks (e.g., RDMA) for efficient model loading between nodes, which can reduce memory footprint while maintaining low-latency model startup. 
}

\section{Related Work}
\label{sec:discuss}

\PHM{\shep{Serverless Inference.}}
\shep{
In addition to the serverless platforms discussed in \S\ref{sec:background-limitations}, there are several recent works on serverless inference.
StreamBox~\cite{wu_streambox_2024} and Dilu~\cite{lv_dilu_2025} support spatial GPU sharing for concurrent model execution; FaaSTube~\cite{wu2024faastube} improves data sharing within inference workflows.
These works are orthogonal to \SysName and can be integrated for enhanced performance.
ServerlessLLM~\cite{fu_serverlessllm_2024} and Medusa~\cite{medusa_asplos25} propose cold-start optimizations specialized for large language models, which is complementary to \SysName and can be used to accelerate those specific models.
}

\PHM{\revise{Host-to-GPU data swapping.}}
\add{
Many other systems have leveraged host-to-GPU data swapping in general deep learning and GPU workloads~\cite{rhu_vdnn_2016,yu_salus_nodate,huang_swapadvisor_2020,xiao_antman_nodate,kim_batch-aware_2020,choi_memory_nodate,gujarati_serving_2020,sheng2023highthroughput}.
For example, vDNN~\cite{rhu_vdnn_2016}, Salus~\cite{yu_salus_nodate} and SwapAdvisor~\cite{huang_swapadvisor_2020} leverage host memory for deep learning jobs with large GPU memory footprints; Batch-aware~\cite{kim_batch-aware_2020} and HUVM~\cite{choi_memory_nodate} optimize GPU memory access for general-purpose workloads; POS~\cite{huang_parallelgpuos_2024} supports efficient GPU checkpointing and restoring.
Compared with \SysName, these systems are not specifically designed for model inference and do not account for its SLO attainment.
Inference systems such as PipeSwitch~\cite{bai_pipeswitch_nodate} and DeepPlan~\cite{deepplan_2023} improve host-to-GPU model loading for fast switching.
Unlike \SysName, these systems require detailed model-specific knowledge and do not target meeting model-level SLOs in a shared, multi-tenant serverless environment.
}


\PHM{\revise{GPU remoting.}}
\revise{
GPU remoting techniques have been employed in different layers for GPU virtualization~\cite{gVirtuS_paper,rcuda,cu2rcu,yu_ava_2020}.
Existing solutions like GVirtuS~\cite{gVirtuS_paper} and rCUDA~\cite{rcuda} primarily focus on general-purpose workloads.
\SysName applies GPU remoting in serverless inference, leveraging its characteristics for asynchronous, low-latency API redirection.}

\PHM{\revise{Spatio-temporal GPU sharing.}}
\revise{
Existing techniques have investigated the spatial and temporal GPU sharing to improve overall utilization~\cite{cgpu,vgpu,mig_gpu,mps_gpu,choi_serving_2022,han_microsecond-scale_2022,strati2024orion,wu_streambox_2024}. 
These techniques are orthogonal to \SysName and can be directly applied, which allow partitioning a physical GPU into multiple virtual instances to late-bind more functions.}
\section{Conclusion}
\label{sec:conclusion}


This paper introduces \SysName, a serverless platform for SLO-aware and
GPU-efficient model inference. \SysName employs a late binding approach,
managing inference functions in host memory and dynamically swapping them to
a pool of GPUs upon request arrivals. This approach enables pay-per-GPU-use
billing and maximizes resource utilization. Additionally, \SysName proposes
request scheduling and model management policies to meet latency SLOs for
inference functions while minimizing resource costs. \SysName has
been beta released in a large commercial serverless platform,
successfully serving up to 465k requests per day and achieving 70\% and 65\%
GPU cost savings for users and the platform, respectively.


\section*{Acknowledgments}

We thank the anonymous reviewers and our shepherd, Somali Chaterji, for their insightful comments that helped improve this work. We also thank Bohui Wu and Zhexiang Zhang for their help in experiments. 
This work was supported in part by the Alibaba Innovative Research (AIR) Grant, RGC CRF Grant (Ref. \#C6015-23G), RGC GRF Grants (Ref. \#16217124 and \#16210822), NSFC/RGC CRS Grant (Ref. \#CRS\_HKUST601/24), and CUHK-Shenzhen Research Grant (UDF01003466).



\bibliographystyle{plain}
\bibliography{references}

\begin{thebibliography}{10}

\bibitem{ali_fc}
{Alibaba Cloud Function Compute}.
\newblock \url{https://www.alibabacloud.com/product/function-compute}.

\bibitem{cgpu}
{Aliyun cGPU}.
\newblock
  \url{https://www.alibabacloud.com/help/en/container-service-for-kubernetes/latest/cgpu-overview}.

\bibitem{fc_billing}
{Aliyun Function Compute Billing Scheme}.
\newblock
  \url{https://www.alibabacloud.com/help/en/function-compute/latest/billing-billing}.

\bibitem{fc_gpu}
{Aliyun Function Compute Instance Types and Modes}.
\newblock
  \url{https://www.alibabacloud.com/help/en/function-compute/latest/instance-types-and-instance-modes}.

\bibitem{sagemaker}
{Amazon SageMaker}.
\newblock \url{https://aws.amazon.com/sagemaker/}.

\bibitem{aws_lambda}
{AWS} {Lambda}.
\newblock \url{https://aws.amazon.com/lambda/}.

\bibitem{aws_provisioned}
{AWS Lambda Provisioned Concurrency}.
\newblock
  \url{https://docs.aws.amazon.com/lambda/latest/dg/provisioned-concurrency.html}.

\bibitem{azurefunc}
{Azure} {Functions}.
\newblock \url{https://azure.microsoft.com/en-us/services/functions/}.

\bibitem{gVirtuS_git}
{GVirtuS}.
\newblock \url{https://github.com/gvirtus/GVirtuS}.

\bibitem{llama2}
{Llama2}.
\newblock \url{https://www.llama.com/llama2}.

\bibitem{llama3}
{Llama3}.
\newblock \url{https://www.llama.com/models/llama-3}.

\bibitem{mem_management_url}
{Memory Management on Modern GPU Architectures.}
\newblock
  \url{https://developer.download.nvidia.com/video/gputechconf/gtc/2019/presentation/s9727-memory-management-on-modern-gpu-architectures.pdf}.

\bibitem{mig_gpu}
{Nvidia Multi-Instance GPU}.
\newblock \url{https://www.nvidia.com/en-us/technologies/multi-instance-gpu/}.

\bibitem{mps_gpu}
{Nvidia Multi-Process Service}.
\newblock \url{https://docs.nvidia.com/deploy/mps/}.

\bibitem{vgpu}
{Nvidia Virtual GPU}.
\newblock \url{https://www.nvidia.com/en-us/data-center/virtual-solutions/}.

\bibitem{qwen}
{Qwen}.
\newblock \url{https://github.com/QwenLM/Qwen}.

\bibitem{resnet}
{ResNet in PyTorch}.
\newblock \url{https://pytorch.org/vision/stable/models/resnet.html}.

\bibitem{sdiff}
{Stable Diffusion}.
\newblock
  \url{https://huggingface.co/stable-diffusion-v1-5/stable-diffusion-v1-5}.

\bibitem{agache_firecracker_2020}
Alexandru Agache, Marc Brooker, Alexandra Iordache, Anthony Liguori, Rolf
  Neugebauer, Phil Piwonka, and Diana-Maria Popa.
\newblock Firecracker: Lightweight virtualization for serverless applications.
\newblock In {\em Proc. {USENIX} NSDI}, 2020.

\bibitem{ali_batch_nodate}
Ahsan Ali, Riccardo Pinciroli, Feng Yan, and Evgenia Smirni.
\newblock {BATCH}: Machine learning inference serving on serverless platforms
  with adaptive batching.
\newblock In {\em Proc. {ACM/IEEE} Supercomputing}, 2020.

\bibitem{topo_gpu_2017}
Marcelo Amaral, Jord\`{a} Polo, David Carrera, Seetharami Seelam, and
  Malgorzata Steinder.
\newblock Topology-aware gpu scheduling for learning workloads in cloud
  environments.
\newblock In {\em Proc. ACM SC}, 2017.

\bibitem{bai_pipeswitch_nodate}
Zhihao Bai, Zhen Zhang, Yibo Zhu, and Xin Jin.
\newblock {PipeSwitch}: Fast pipelined context switching for deep learning
  applications.
\newblock In {\em Proc. {USENIX} OSDI}, 2020.

\bibitem{choi_memory_nodate}
Sangjin Choi, Taeksoo Kim, Jinwoo Jeong, Myeongjae Jeon, Youngjin Kwon, Rachata
  Ausavarungnirun, and Jeongseob Ahn.
\newblock Memory harvesting in multi-{GPU} systems with hierarchical unified
  virtual memory.
\newblock In {\em Proc. {USENIX} ATC}, 2022.

\bibitem{choi_serving_2022}
Seungbeom Choi, Sunho Lee, Yeonjae Kim, Jongse Park, Youngjin Kwon, and Jaehyuk
  Huh.
\newblock Serving heterogeneous machine learning models on multi-{GPU} servers
  with spatio-temporal sharing.
\newblock In {\em Proc. {USENIX} ATC}, 2022.

\bibitem{crankshaw_clipper}
Daniel Crankshaw, Xin Wang, Giulio Zhou, Michael~J Franklin, Joseph~E Gonzalez,
  and Ion Stoica.
\newblock Clipper: A low-latency online prediction serving system.
\newblock In {\em Proc. {USENIX} NSDI}, 2017.

\bibitem{devlin2019bert}
Jacob Devlin, Ming-Wei Chang, Kenton Lee, and Kristina Toutanova.
\newblock Bert: Pre-training of deep bidirectional transformers for language
  understanding.
\newblock {\em arXiv preprint arXiv:1810.04805}, 2018.

\bibitem{heter_serverless_2022}
Dong Du, Qingyuan Liu, Xueqiang Jiang, Yubin Xia, Binyu Zang, and Haibo Chen.
\newblock Serverless computing on heterogeneous computers.
\newblock In {\em Proc. {ACM} ASPLOS}, 2022.

\bibitem{rcuda}
José Duato, Antonio~J. Peña, Federico Silla, Rafael Mayo, and Enrique~S.
  Quintana-Ortí.
\newblock rcuda: Reducing the number of gpu-based accelerators in high
  performance clusters.
\newblock In {\em Proc. {IEEE} HPCS}, 2010.

\bibitem{dgsf_2022}
Henrique Fingler, Zhiting Zhu, Esther Yoon, Zhipeng Jia, Emmett Witchel, and
  Christopher~J. Rossbach.
\newblock Dgsf: Disaggregated gpus for serverless functions.
\newblock In {\em Proc. IEEE IPDPS}, 2022.

\bibitem{fu_serverlessllm_2024}
Yao Fu, Leyang Xue, Yeqi Huang, Andrei-Octavian Brabete, Dmitrii Ustiugov,
  Yuvraj Patel, and Luo Mai.
\newblock {ServerlessLLM}: Locality-enhanced serverless inference for large
  language models.
\newblock In {\em Proc. {USENIX} OSDI}, 2024.

\bibitem{gVirtuS_paper}
Giulio Giunta, Raffaele Montella, Giuseppe Agrillo, and Giuseppe Coviello.
\newblock A gpgpu transparent virtualization component for high performance
  computing clouds.
\newblock In {\em Proc. Euro-Par}, 2010.

\bibitem{gujarati_serving_2020}
Arpan Gujarati, Reza Karimi, Safya Alzayat, Wei Hao, Antoine Kaufmann, Ymir
  Vigfusson, and Jonathan Mace.
\newblock Serving {DNNs} like clockwork: Performance predictability from the
  bottom up.
\newblock In {\em Proc. {USENIX} OSDI}, 2020.

\bibitem{han_microsecond-scale_2022}
Mingcong Han, Hanze Zhang, Rong Chen, and Haibo Chen.
\newblock Microsecond-scale preemption for concurrent {GPU}-accelerated {DNN}
  inferences.
\newblock In {\em Proc. {USENIX} OSDI}, 2022.

\bibitem{huang_swapadvisor_2020}
Chien-Chin Huang, Gu~Jin, and Jinyang Li.
\newblock {SwapAdvisor}: Pushing deep learning beyond the {GPU} memory limit
  via smart swapping.
\newblock In {\em Proc. {ACM} ASPLOS}, 2020.

\bibitem{huang_parallelgpuos_2024}
Zhuobin Huang, Xingda Wei, Yingyi Hao, Rong Chen, Mingcong Han, Jinyu Gu, and
  Haibo Chen.
\newblock {PARALLELGPUOS}: A concurrent {OS}-level {GPU} checkpoint and restore
  system using validated speculation.
\newblock {\em arXiv preprint arXiv:2405.12079}, 2024.

\bibitem{deepplan_2023}
Jinwoo Jeong, Seungsu Baek, and Jeongseob Ahn.
\newblock Fast and efficient model serving using multi-gpus with
  direct-host-access.
\newblock In {\em Proc. ACM EuroSys}, 2023.

\bibitem{jia_nightcore_2021}
Zhipeng Jia and Emmett Witchel.
\newblock Nightcore: Efficient and scalable serverless computing for
  latency-sensitive, interactive microservices.
\newblock In {\em Proc. {ACM} ASPLOS}, 2021.

\bibitem{jiang2020unified}
Yimin Jiang, Yibo Zhu, Chang Lan, Bairen Yi, Yong Cui, and Chuanxiong Guo.
\newblock A unified architecture for accelerating distributed dnn training in
  heterogeneous gpu/cpu clusters.
\newblock In {\em Proc. {USENIX} OSDI}, 2020.

\bibitem{kim_batch-aware_2020}
Hyojong Kim, Jaewoong Sim, Prasun Gera, Ramyad Hadidi, and Hyesoon Kim.
\newblock Batch-aware unified memory management in {GPUs} for irregular
  workloads.
\newblock In {\em Proc. {ACM} ASPLOS}, 2020.

\bibitem{buddy_memory}
Kenneth~C. Knowlton.
\newblock A fast storage allocator.
\newblock {\em Commun. ACM}, 8(10):623--624, 1965.

\bibitem{kosaian_parity_2019}
Jack Kosaian, K.~V. Rashmi, and Shivaram Venkataraman.
\newblock Parity models: erasure-coded resilience for prediction serving
  systems.
\newblock In {\em Proc. ACM SOSP}, 2019.

\bibitem{lee_pretzel}
Yunseong Lee, Alberto Scolari, Byung-Gon Chun, Marco~Domenico Santambrogio,
  Markus Weimer, and Matteo Interlandi.
\newblock {PRETZEL}: Opening the black box of machine learning prediction
  serving systems.
\newblock In {\em Proc. {USENIX} OSDI}, 2018.

\bibitem{lv_dilu_2025}
Cunchi Lv, Xiao Shi, Zhengyu Lei, Jinyue Huang, Wenting Tan, Xiaohui Zheng, and
  Xiaofang Zhao.
\newblock Dilu: Enabling {GPU} resourcing-on-demand for serverless {DL} serving
  via introspective elasticity.
\newblock In {\em Proc. {ACM} ASPLOS}, 2025.

\bibitem{peng_capuchin_2020}
Xuan Peng, Xuanhua Shi, Hulin Dai, Hai Jin, Weiliang Ma, Qian Xiong, Fan Yang,
  and Xuehai Qian.
\newblock Capuchin: Tensor-based {GPU} memory management for deep learning.
\newblock In {\em Proc. {ACM} ASPLOS}, 2020.

\bibitem{cu2rcu}
C.~Reaño, A.~J. Peña, F.~Silla, J.~Duato, R.~Mayo, and E.~S. Quintana-Ortí.
\newblock Cu2rcu: Towards the complete rcuda remote gpu virtualization and
  sharing solution.
\newblock In {\em Proc. {IEEE} HiPC}, 2012.

\bibitem{rhu_vdnn_2016}
Minsoo Rhu, Natalia Gimelshein, Jason Clemons, Arslan Zulfiqar, and Stephen~W.
  Keckler.
\newblock {vDNN}: Virtualized deep neural networks for scalable,
  memory-efficient neural network design.
\newblock In {\em Proc. {ACM/IEEE} MICRO}, 2016.

\bibitem{romero_infaas_nodate}
Francisco Romero, Qian Li, Neeraja~J Yadwadkar, and Christos Kozyrakis.
\newblock {INFaaS}: Automated model-less inference serving.
\newblock In {\em Proc. {USENIX} ATC}, 2021.

\bibitem{shahrad_serverless_2020}
Mohammad Shahrad, Rodrigo Fonseca, I{\~{n}}igo Goiri, Gohar Chaudhry, Paul
  Batum, Jason Cooke, Eduardo Laureano, Colby Tresness, Mark Russinovich, and
  Ricardo Bianchini.
\newblock Serverless in the wild: Characterizing and optimizing the serverless
  workload at a large cloud provider.
\newblock In {\em Proc. {USENIX} {ATC}}, 2020.

\bibitem{shen_nexus_2019}
Haichen Shen, Lequn Chen, Yuchen Jin, Liangyu Zhao, Bingyu Kong, Matthai
  Philipose, Arvind Krishnamurthy, and Ravi Sundaram.
\newblock Nexus: a {GPU} cluster engine for accelerating {DNN}-based video
  analysis.
\newblock In {\em Proc. {ACM} SOSP}, 2019.

\bibitem{sheng2023highthroughput}
Ying Sheng, Lianmin Zheng, Binhang Yuan, Zhuohan Li, Max Ryabinin, Daniel~Y.
  Fu, Zhiqiang Xie, Beidi Chen, Clark Barrett, Joseph~E. Gonzalez, Percy Liang,
  Christopher Ré, Ion Stoica, and Ce~Zhang.
\newblock High-throughput generative inference of large language models with a
  single gpu.
\newblock {\em arXiv preprint arXiv:2303.06865}, 2023.

\bibitem{strati2024orion}
Foteini Strati, Xianzhe Ma, and Ana Klimovic.
\newblock Orion: Interference-aware, fine-grained gpu sharing for ml
  applications.
\newblock In {\em Proc. ACM EuroSys}, 2024.

\bibitem{tian2022Owl}
Huangshi Tian, Suyi Li, Ao~Wang, Wei Wang, Tianlong Wu, and Haoran Yang.
\newblock Owl: Performance-aware scheduling for resource-efficient
  function-as-a-service cloud.
\newblock In {\em Proc. ACM SoCC}, 2022.

\bibitem{ustiugov_benchmarking_2021}
Dmitrii Ustiugov, Plamen Petrov, Marios Kogias, Edouard Bugnion, and Boris
  Grot.
\newblock Benchmarking, analysis, and optimization of serverless function
  snapshots.
\newblock In {\em Proc. {ACM} ASPLOS}, 2021.

\bibitem{wang_faasnet_nodate}
Ao~Wang, Shuai Chang, Huangshi Tian, Hongqi Wang, Haoran Yang, Huiba Li, Rui
  Du, and Yue Cheng.
\newblock {FaaSNet}: Scalable and fast provisioning of custom serverless
  container runtimes at alibaba cloud function compute.
\newblock In {\em Proc. {USENIX} ATC}, 2021.

\bibitem{wu2024faastube}
Hao Wu, Junxiao Deng, Minchen Yu, Yue Yu, Yaochen Liu, Hao Fan, Song Wu, and
  Wei Wang.
\newblock Faastube: Optimizing gpu-oriented data transfer for serverless
  computing.
\newblock {\em arXiv preprint arXiv:2411.01830}, 2024.

\bibitem{wu_streambox_2024}
Hao Wu, Yue Yu, Junxiao Deng, Shadi Ibrahim, Song Wu, Hao Fan, Ziyue Cheng, and
  Hai Jin.
\newblock {StreamBox}: A lightweight {GPU} {SandBox} for serverless inference
  workflow.
\newblock In {\em Proc. {USENIX} ATC}, 2024.

\bibitem{xiao_antman_nodate}
Wencong Xiao, Shiru Ren, Yong Li, Yang Zhang, Pengyang Hou, Zhi Li, Yihui Feng,
  Wei Lin, and Yangqing Jia.
\newblock {AntMan}: Dynamic scaling on {GPU} clusters for deep learning.
\newblock In {\em Proc. {USENIX} OSDI}, 2020.

\bibitem{yang_infless_2022}
Yanan Yang, Laiping Zhao, Yiming Li, Huanyu Zhang, Jie Li, Mingyang Zhao,
  Xingzhen Chen, and Keqiu Li.
\newblock {INFless}: a native serverless system for low-latency,
  high-throughput inference.
\newblock In {\em Proc. ACM ASPLOS}, 2022.

\bibitem{yu_ava_2020}
Hangchen Yu, Arthur~Michener Peters, Amogh Akshintala, and Christopher~J.
  Rossbach.
\newblock {AvA}: Accelerated virtualization of accelerators.
\newblock In {\em Proc. {ACM} ASPLOS}, 2020.

\bibitem{pheromone}
Minchen Yu, Tingjia Cao, Wei Wang, and Ruichuan Chen.
\newblock Following the data, not the function: Rethinking function
  orchestration in serverless computing.
\newblock In {\em Proc. {USENIX} NSDI}, 2023.

\bibitem{yu_gillis_icdcs}
Minchen Yu, Zhifeng Jiang, Hok~Chun Ng, Wei Wang, Ruichuan Chen, and Bo~Li.
\newblock Gillis: Serving large neural networks in serverless functions with
  automatic model partitioning.
\newblock In {\em Proc. IEEE ICDCS}, 2021.

\bibitem{yu_salus_nodate}
Peifeng Yu and Mosharaf Chowdhury.
\newblock Salus: Fine-grained {GPU} sharing primitives for deep learning
  applications.
\newblock In {\em Proc. MLSys}, 2020.

\bibitem{medusa_asplos25}
Shaoxun Zeng, Minhui Xie, Shiwei Gao, Youmin Chen, and Youyou Lu.
\newblock Medusa: Accelerating serverless {LLM} inference with materialization.
\newblock In {\em Proc. {ACM} ASPLOS}, 2025.

\bibitem{zhang_mark:_2019}
Chengliang Zhang, Minchen Yu, Wei Wang, and Feng Yan.
\newblock {MArk}: Exploiting cloud services for cost-effective, {SLO}-aware
  machine learning inference serving.
\newblock In {\em Proc. USENIX ATC}, 2019.

\bibitem{zhang_caerus_nodate}
Hong Zhang, Yupeng Tang, Anurag Khandelwal, Jingrong Chen, and Ion Stoica.
\newblock Caerus: {NIMBLE} task scheduling for serverless analytics.
\newblock In {\em Proc. {USENIX} NSDI}, 2021.

\bibitem{zhang_shepherd_nodate}
Hong Zhang, Yupeng Tang, Anurag Khandelwal, and Ion Stoica.
\newblock {SHEPHERD}: Serving {DNNs} in the wild.
\newblock In {\em Proc. {USENIX} NSDI}, 2023.

\bibitem{zhong_distserve_2024}
Yinmin Zhong, Shengyu Liu, Junda Chen, Jianbo Hu, Yibo Zhu, Xuanzhe Liu, Xin
  Jin, and Hao Zhang.
\newblock {DistServe}: Disaggregating prefill and decoding for
  goodput-optimized large language model serving.
\newblock In {\em 18th {USENIX} Symposium on Operating Systems Design and
  Implementation ({OSDI} 24)}, 2024.

\end{thebibliography}
\newpage

\newpage
\appendix
\section{Appendix}

\subsection{CUDA APIs and Batch-Level API Redirection}
\label{sec:appendix_cuda_api}

\SysName performs asynchronous CUDA API redirection to reduce communication overhead for efficient GPU remoting (see \S\ref{sec:system_remote}).
We divide CUDA APIs into two categories, i.e., asynchronous and synchronous APIs, according to whether they require GPU-to-host data transfer and update state in host.
Table~\ref{tab:cuda_api} lists the primary CUDA APIs supported in \SysName and their categories.
CUDA APIs issued by intermediate steps during model inference are generally asynchronous.
In addition to listed APIs, model inference can also trigger a few other CUDA APIs in our experiments, e.g., \texttt{cudaGetDevice}.
These APIs do not affect inference execution and thus \SysName can cache their results in GPU clients without repeatedly querying the executor, which further reduces communications.

\SysName enhances the efficiency of asynchronous API redirection by bundling multiple consecutive CUDA API calls into a single batch for joint transimission.
Determining the ideal batch size is crucial to balance communication costs and the waiting time: larger batches decrease communication overhead, but they necessitate a longer period to gather sufficient calls; conversely, smaller batches reduce waiting delays but increase communication costs due to more frequent transmissions.
\SysName develops an effective batching strategy by leveraging the repetitive nature of API call patterns across various models, which often use common building blocks (e.g., convolutional layers).
Therefore we can conduct profiling of common API call sequences under various batch sizes to determine an effective size that consistently yields good performance.

\begin{table}[t]
    \centering
    \caption{Primary CUDA APIs supported in \SysName, which we divide into asynchronous and synchronous APIs according to their semantics.}
    \footnotesize
    \label{tab:cuda_api}
    \begin{tabular}{lll}
        \toprule
        \textbf{CUDA library} & \textbf{API}  \\
        \midrule
        CRT API (Async.) & \texttt{cudaMemcpyAsync} \\
        & \texttt{cudaMemsetAsync} \\
        & \texttt{cudaLaunchKernel} \\
        & \texttt{cudaFree} \\

        CRT API (Sync.) & \texttt{cudaMalloc} \\
        & \texttt{cudaMemcpy} \\        
        & \texttt{cudaStreamCreate} \\        
        & \texttt{cudaStreamCreateWithFlags} \\        
        & \texttt{cudaStreamCreateWithPriority} \\        
        & \texttt{cudaStreamSynchronize} \\        
        & \texttt{cudaEventCreateWithFlags} \\        
        & \texttt{cudaEventQuery} \\        
        & \texttt{cudaGetDeviceCount} \\       
        & \texttt{cudaGetDeviceProperties} \\       
        & \texttt{cudaDeviceSynchronize} \\        

        \midrule
        \texttt{cuDNN} (Async.) & \texttt{cudnnSetStream} \\
        & \texttt{cudnnCreateFilterDescriptor} \\        
        & \texttt{cudnnSetFilterNdDescriptor} \\        
        & \texttt{cudnnDestroyFilterDescriptor} \\        
        & \texttt{cudnnCreateConvolutionDescriptor} \\        
        & \texttt{cudnnSetConvolutionGroupCount} \\        
        & \texttt{cudnnSetConvolutionNdDescriptor} \\        
        & \texttt{cudnnSetConvolutionMathType} \\        
        & \texttt{cudnnDestroyConvolutionDescriptor} \\        
        & \texttt{cudnnCreateTensorDescriptor} \\        
        & \texttt{cudnnSetTensorNdDescriptor} \\        
        & \texttt{cudnnDestroyTensorDescriptor} \\        
        & \texttt{cudnnConvolutionForward} \\        
        & \texttt{cudnnBatchNormalizationForwardInference} \\   
        
        \texttt{cuDNN} (Sync.) & \texttt{cudnnCreate} \\
        & \texttt{cudnnGetConvolutionForwardAlgorithm\_v7} \\
        
        \midrule
        
        \texttt{cuBLAS} (Async.) & \texttt{cublasSetStream} \\
        & \texttt{cublasSetMathMode} \\        
        & \texttt{cublasSgemm} \\        
        & \texttt{cublasSgemmStridedBatched} \\        

        \texttt{cuBLAS} (Sync.) & \texttt{cublasCreate} \\
        & \texttt{cublasGetMathMode} \\

        \bottomrule
    \end{tabular}
\end{table}

\subsection{Memory Block Allocation}
\label{sec:appendix_block_alloc}

\SysName extends the Buddy memory allocation algorithm by leverages two characteristics of inference, effectively reducing memory fragmentation.
First, we leverage the block patterns of ML frameworks to improve memory sharing.
We observe that ML frameworks like PyTorch typically use fixed-sized memory blocks (e.g., 20MB) to consolidate small- and moderate-sized data, resulting in high popularity and easy sharing of these blocks across various models.
Motivated by this observation, \SysName proposes a two-tier block management scheme.
\SysName divides all GPU memory into a number of \emph{memory partitions}, where each partition can either host common fixed-sized blocks or manage irregular-sized blocks via the Buddy allocation algorithm.
When functions request memory blocks with common sizes, \SysName consolidates them on associated memory partitions, effectively reducing the internal memory fragmentation and enabling efficient sharing of those partitions.

Second, we note that memory blocks of a single model are usually accessed in their entirety during model swapping and execution, and are reclaimed together after model eviction.
This observation inspires \SysName to package blocks from the same model as tightly as possible, such as collocating them into a single memory partition.
By employing this approach, model eviction can also free the entire memory partitions, making them available for future block allocation.
Furthermore, \SysName periodically consolidates blocks across partitions to reduce memory fragments.

\subsection{Function Request Prioritization}
\label{sec:appendix_auto_config}

With RRCs, \SysName can divide functions into high- and low-priority groups, and then determine the request execution order (see \S\ref{sec:algo_queueing}).
Determining the RRC boundary between the two groups can be challenging --- having too many (few) high-priority functions can be too aggressive (conservative) to enable more SLO-compliant requests.
In \SysName, we use a threshold $\alpha \in [0, 1]$ to indicate the boundary and determine how functions are prioritized:
we prioritize more functions by increasing $\alpha$; when $\alpha$ is 1, all functions are put in the high-priority group.
In particular, consider a node with $N$ functions sorted by RRCs, and let $RRC_i$ be the RRC of function $i$.
We put the first sorted $k$ functions in the high-priority group, where $k$ is the largest integer such that $\sum_{j=1}^{k} \max (RRC_j, 0) \leq \alpha \cdot \sum_{i=1}^{N} \max (RRC_i, 0)$.
\SysName automatically configures $\alpha$ at runtime based on the overall load and function SLOs.
When there is a load surge and the number of SLO-compliant functions decreases, \SysName turns more conservative with a small $\alpha$; otherwise, $\alpha$ increases to prioritize more functions.

\SysName can automatically configure $\alpha$ based on overall load such as to maximize the number of SLO-compliant functions per node.
Intuitively, when the load is low and an increasing number of functions can satisfy SLOs, $\alpha$ should grow to prioritize more functions to enable more SLO-compliant functions.
On the contrary, \SysName should be conservative to prevent functions to violate their SLOs by decreasing $\alpha$ when a node is overloaded.
Therefore, we propose an auto-configuration algorithm for $\alpha$, which is inspired by TCP congestion control.
Algorithm~\ref{algo:alpha-config} shows the pseudo code, where $scalar$ and $threshold$ are two parameters to determine how much and when $\alpha$ should change.
We by default set $scalar$ to 2 and $threshold$ to 0.04, which is able to properly adjust $\alpha$ according to our profiling.

\begin{algorithm}[tb]
	\caption{$\alpha$ Auto-configuration.}
    \label{algo:alpha-config}
    \footnotesize
    \begin{algorithmic}[1]
        \State{$scalar$ --- scale factor larger than 1 that controls the rate of $\alpha$ change}
        \State{$threshold$ --- threshold to trigger $\alpha$ change}
        \Function{PeriodicConfig}{}
            \State{$\alpha \gets $ $\alpha$ in last period} 
            \State{$last\_ratio \gets $ ratio of SLO-compliant functions in the last period}
            \State{$new\_ratio \gets $ ratio of SLO-compliant functions in this period}

            \If{$new\_ratio -  last\_ratio > |threshold|$}
                \State{$\alpha \gets \min(\alpha \cdot scalar, 1)$} \Comment{Increase $\alpha$}
            \ElsIf{$new\_ratio -  last\_ratio < -|threshold|$}
                \State{$\alpha \gets \alpha / scalar$} \Comment{Decrease $\alpha$}
            \Else
                \State{$\alpha \gets \alpha$} \Comment{Keep $\alpha$ unchanged}
            \EndIf
        \EndFunction

	\end{algorithmic}
\end{algorithm}

The request execution order within each priority group is also determined using function RRCs.  We maintain two queues for the two priority groups.
In the high-priority queue, requests are prioritized in a \emph{reverse} order of RRCs.
This strategy favors functions with a slightly lower but still high likelihood of SLO compliance (i.e., with small positive RRCs) over those already meeting SLOs (i.e., with non-positive RRCs).  This effectively increases the number of SLO-compliant requests whenever feasible.
In contrast, requests in the low-priority queue are executed following the order of RRCs.

\begin{figure}
    \centering
    \includegraphics[width=0.47\textwidth]{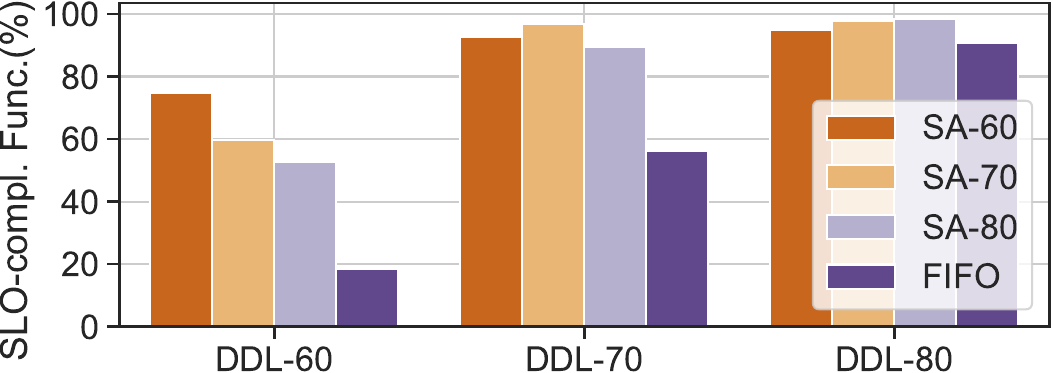}
    \caption{Ratio of SLO-compliant functions using FIFO and \SysName's SLO-aware (SA) policies. We vary deadlines from 60 ms to 80 ms, and vary the objectives in SA accordingly.}
    \label{fig:slo_aware}
	\vspace{-.1in}
\end{figure}

\PHM{Experimental results.}
We evaluate \SysName's SLO-aware request queueing policy by compare \SysName with \SysName-FIFO (see \S\ref{sec:eval_node}) under 560 ResNet-152 functions and vary their deadlines from 60 ms to 80 ms.
Fig.~\ref{fig:slo_aware} shows the ratio of SLO-compliant functions using the FIFO policy and \SysName's SLO-aware (SA) policy that can adjust the request execution order based on SLO objectives.  We set the deadlines to 60 ms, 70 ms, and 80 ms, respectively, denoted as SA-60, SA-70, and SA-80.
All of them outperform \SysName-FIFO with either deadline.

\end{document}